\documentclass[twocolumn]{aastex631}

\usepackage{newtxtext,newtxmath}
\usepackage[T1]{fontenc}

\usepackage{graphicx}	
\usepackage{amsmath}	

\usepackage{graphicx}
\usepackage{amsmath}
\usepackage{subfigure}
\usepackage{multirow}
\usepackage{threeparttable}
\usepackage{array}
\usepackage{natbib}
\usepackage{xcolor}
\usepackage{hyperref}
\usepackage[version=4]{mhchem}
\usepackage{bm}
\usepackage{ulem}

\newcommand{\msun}{\mbox{${\rm M}_\odot$}}
\newcommand{\mstar}{\mbox{${M}_{\rm star}$}}
\newcommand{\cmjj}{\mbox{${\rm cm^{-2}}$}}

\newcommand{\lya}{\mbox{${\rm Ly}\alpha$}}
\newcommand{\lyb}{\mbox{${\rm Ly}\beta$}}
\newcommand{\kms}{\,{\rm{km\,s}^{-1}}}
\newcommand{\cc}{{\rm cm^{-3}}}
\newcommand{\dproj}{\mbox{$d_{\rm proj}$}}
\newcommand{\rvir}{\mbox{$r_{\rm vir}$}}
\newcommand{\dvir}{\mbox{$\dproj/r_{\rm vir}$}}
\newcommand{\neo}{${N_{\rm NeVIII}/N_{\rm OVI}}$}

\defcitealias{Johnson:2017aa}{J17}
\defcitealias{CUBSVI}{CUBS VI}
\defcitealias{CUBSV}{CUBS V}
\defcitealias{Cooper:2021aa}{CUBS IV}
\defcitealias{Zahedy:2021aa}{CUBS III}
\defcitealias{Stern:2018aa}{S18}
\defcitealias{Qu:2018aa}{QB18}

\begin{document}

\title[CUBS VII: \ion{O}{6} and \ion{Ne}{8} at $z\approx 0.5$]{The Cosmic Ultraviolet Baryon Survey (CUBS) VII: on the warm-hot circumgalactic medium probed by \ion{O}{6} and \ion{Ne}{8} at $\bm{0.4 \lesssim z\lesssim0.7}$}

\correspondingauthor{Zhijie Qu}
\email{quzhijie@uchicago.edu}

\author{Zhijie Qu}
\affiliation{Department of Astronomy \& Astrophysics, The University of Chicago, 5640 S. Ellis Ave., Chicago, IL 60637, USA}

\author{Hsiao-Wen Chen}
\affiliation{Department of Astronomy \& Astrophysics, The University of Chicago, 5640 S. Ellis Ave., Chicago, IL 60637, USA}

\author{Sean D. Johnson}
\affiliation{Department of Astronomy, University of Michigan, Ann Arbor, MI 48109, USA}
\author{Gwen C. Rudie}
\affiliation{The Observatories of the Carnegie Institution for Science, 813 Santa Barbara Street, Pasadena, CA 91101, USA}
\author{Fakhri S. Zahedy}
\affiliation{The Observatories of the Carnegie Institution for Science, 813 Santa Barbara Street, Pasadena, CA 91101, USA}
\author{David DePalma}
\affiliation{Department of Physics, 77 Massachusetts Avenue, Cambridge, Massachusetts 02139, USA
}
\affiliation{MIT-Kavli Institute for Astrophysics and Space Research, 77 Massachusetts Ave., Cambridge, MA 02139, USA}
\author{Joop Schaye}
\affiliation{Leiden Observatory, Leiden University, PO Box 9513, NL-2300 RA Leiden, the Netherlands}
\author{Erin T. Boettcher}
\affiliation{Department of Astronomy, University of Maryland, College Park, MD 20742, USA}
\affiliation{X-ray Astrophysics Laboratory, NASA/GSFC, Greenbelt, MD 20771, USA}
\affiliation{Center for Research and Exploration in Space Science and Technology, NASA/GSFC, Greenbelt, MD 20771, USA}
\author{Sebastiano Cantalupo}
\affiliation{Department of Physics, University of Milan Bicocca, Piazza della Scienza 3, I-20126 Milano, Italy}
\author{Mandy C. Chen}
\affiliation{Department of Astronomy \& Astrophysics, The University of Chicago, 5640 S. Ellis Ave., Chicago, IL 60637, USA}
\author{Claude-Andr\'e Faucher-Gigu\`ere}
\affiliation{Department of Physics \& Astronomy, Center for Interdisciplinary Exploration and Research in Astrophysics (CIERA), Northwestern University, 1800 Sherman Avenue, Evanston, IL 60201, USA }
\author{Jennifer I-Hsiu Li}
\affiliation{Department of Astronomy, University of Michigan, Ann Arbor, MI 48109, USA}
\affiliation{Michigan Institute for Data Science, University of Michigan, Ann Arbor, MI, 48109, USA}
\author{John S. Mulchaey}
\affiliation{The Observatories of the Carnegie Institution for Science, 813 Santa Barbara Street, Pasadena, CA 91101, USA}
\author{Patrick Petitjean}
\affiliation{Institut d'Astrophysique de Paris, CNRS-SU, UMR 7095, 98bis bd Arago, Paris F-75014, France}
\author{Marc Rafelski}
\affiliation{Space Telescope Science Institute, Baltimore, MD 21218, USA}
\affiliation{Department of Physics \& Astronomy, Johns Hopkins University, Baltimore, MD 21218, USA}



\begin{abstract}
This paper presents a newly established sample of 103 unique galaxies or galaxy groups at $0.4\lesssim z\lesssim 0.7$ from the Cosmic Ultraviolet Baryon Survey (CUBS) for studying the warm-hot circumgalactic medium (CGM) probed by both \ion{O}{6} and \ion{Ne}{8} absorption.
The galaxies and associated neighbors are identified at $< 1$ physical Mpc from the sightlines toward 15 CUBS QSOs at $z_{\rm QSO}\gtrsim 0.8$.
A total of 30 galaxies or galaxy groups exhibit associated \ion{O}{6} $\lambda\lambda$ 1031, 1037 doublet absorption within a line-of-sight velocity interval of $\pm 250 \kms$, while the rest show no trace of \ion{O}{6} to a detection limit of $\log N_{\rm OVI}/\cmjj\approx 13.7$.
Meanwhile, only five galaxies or galaxy groups exhibit the \ion{Ne}{8} $\lambda\lambda$ 770, 780 doublet absorption, down to a limiting column density of $\log N_{\rm NeVIII}/\cmjj\approx 14.0$.
These \ion{O}{6}- and \ion{Ne}{8}-bearing halos reside in different galaxy environments with stellar masses ranging from $\log \mstar/\msun \approx 8$ to $\approx11.5$.
The warm-hot CGM around galaxies of different stellar masses and star formation rates exhibits different spatial profiles and kinematics.
In particular, star-forming galaxies with $\log \mstar/\msun\approx 9-11$ show a significant concentration of metal-enriched warm-hot CGM within the virial radius, while massive quiescent galaxies exhibit flatter radial profiles of both column densities and covering fractions.
In addition, the velocity dispersion of \ion{O}{6} absorption is broad with $\sigma_\varv > 40\kms$ for galaxies of $\log \mstar/\msun > 9$ within the virial radius, suggesting a more dynamic warm-hot halo around these galaxies.
Finally, the warm-hot CGM probed by \ion{O}{6} and \ion{Ne}{8} is suggested to be the dominant phase in sub-$L^*$ galaxies with $\log \mstar/\msun\approx9-10$ based on their high ionization fractions in the CGM.
\end{abstract}





\section{Introduction}

The circumgalactic medium (CGM) is a multiphase gaseous reservoir surrounding galaxies.  It spans a range in temperature and ionization conditions from the cool photoionized gas ($\log T/{\rm K}\approx4$) to the collisionally ionized hot gas ($\log T/{\rm K}\approx6$; see \citealt{Donahue:2022aa, Faucher-Giguere:2023aa} for recent reviews).
In this multiphase gas, the warm-hot phase at $\log T/{\rm K}\sim 5-6$ is of particular importance in understanding the transformation of the CGM and its connection with galaxy evolution (e.g., \citealt{Tepper:2013aa, Oppenheimer:2016aa, Rahmati:2016aa, Nelson:2018aa, McQuinn:2018aa, Stern:2019aa, Faerman:2020aa, Wijers:2020aa, Ho:2021aa, Appleby:2023aa, Wijers:2024aa}).
The warm-hot gas may be short-lived in the CGM, compared to the cool and hot phases because of its high radiative emissivity \citep[e.g.,][]{Oppenheimer:2013aa, Gnat:2017aa}.
At the same time, it may be crucial for the survival or growth of cool clouds as an interface mixed in the hot gas \citep[e.g.,][]{Gronke:2022aa}, making the warm-hot CGM potential fuel for future star formation in galaxies.

In observation, QSO spectroscopy is a powerful tool for characterizing the warm-hot gas by measuring highly ionized species (e.g., \ion{O}{6} and \ion{Ne}{8}) in either local Universe  \citep[e.g.,][]{York:1974aa, Sembach:2003aa, Lehner:2011aa}, at $z\approx0.1-1$ \citep[e.g.,][]{Savage:2005aa, Danforth:2008aa, Thom:2008ab, Tripp:2008aa, Mulchaey:2009aa, Narayanan:2012aa, Meiring:2013aa, Stocke:2014aa, Johnson:2015aa, Danforth:2016aa, Qu:2016aa, Werk:2016aa, Burchett:2019aa, Sankar:2020aa, Tchernyshyov:2022aa} or $z\gtrsim 2$ \citep[e.g.,][]{Bergeron:2002aa, Simcoe:2004aa, Muzahid:2012aa, Rudie:2019aa}.
In particular, \ion{O}{6} and \ion{Ne}{8} are formed with respective ionization potentials of 113.9 eV and 207.3 eV, and destroyed at 138.2 eV and 239.1 eV. 
Under the assumption of collisional ionization equilibrium (CIE), \ion{O}{6} and \ion{Ne}{8} exhibit ionization temperatures of $\log T/{\rm K} \approx 5.5$ and $5.8$, respectively.

In principle, \ion{O}{6} and \ion{Ne}{8} may also be photoionized by the ultraviolet background (UVB; e.g., \citealt{Haardt:2001aa, Hussain:2017aa, Khaire:2019aa, Faucher-Giguere:2020aa}) in the density range of $\log n_{\rm H}/{\rm cm^{-3}} \approx -6$ to $-4$.
In practice, the low density of $\log n_{\rm H}/{\rm cm^{-3}}\lesssim -5$ is rarely seen in the CGM in simulations \citep[e.g.][]{Rahmati:2016aa}.
The expected low gas density implies a temperature of $\log T/{\rm K}> 4.5$ under photoionization equilibrium (PIE).
Furthermore, careful comparison of line profiles of \ion{O}{6} and \ion{Ne}{8} with low ions (e.g., \ion{O}{2} and \ion{C}{2}; tracing cool photoionized CGM of $\log T/{\rm K}\lesssim 4.5$) suggests that they are not in the same phase \citep[e.g.,][]{Zahedy:2019aa, Rudie:2019aa,  Cooper:2021aa, Sameer:2021aa}.
Therefore, the \ion{O}{6} and \ion{Ne}{8} absorption transitions trace the warm-hot gas at $\log T/{\rm K} > 4.5$ in the CGM, irrespective of the exact ionization mechanism.

In practice, \ion{O}{6} and \ion{Ne}{8} are detected as strong doublet transitions at $1031, 1037$\,\AA~ and $770, 780$\,\AA~, respectively, in the far-ultraviolet (FUV) band, which enable robust identifications of these highly ionized species based on the anticipated wavelength separation and doublet ratio.
In addition, both oxygen and neon are $\alpha$-elements with roughly constant relative abundance [Ne/O] and high abundances.  Combining empirical constraints of these two ions, therefore, provides a sensitive probe of the physical conditions of the warm-hot CGM.

In the past few years, deep galaxy surveys have been performed in fields of distant UV bright QSOs with high-quality FUV spectra available for investigating the connection between galaxies and the CGM up to $z\approx 0.5-1.0$ \citep[e.g.,][]{Fossati:2019aa, Chen:2020aa, Dutta:2020aa, Lofthouse:2020aa, Muzahid:2021aa, Wilde:2021aa}, with the warm-hot gas being a significant component in these efforts \citep[e.g.,][]{Chen:2009aa, Johnson:2015aa, Burchett:2019aa, Tchernyshyov:2022aa, Tchernyshyov:2023aa}.
As part of the Cosmic Ultraviolet Baryon Survey (CUBS) program (\citealt{Chen:2020aa}), we are constructing a sample of galaxies from dwarfs with stellar mass $\log\,\mstar/\msun<9$ to massive quiescent galaxies of $\log\,\mstar/\msun>11$ in the redshift range of $z\approx 0.4-0.7$.
All galaxies are selected based on their proximity to the QSO line of sight without any prior information on absorption features, leading to an unbiased sample for absorption property analysis.
The redshift range of $z\approx 0.4-0.7$ is chosen to cover both the \ion{O}{6} and \ion{Ne}{8} doublets in the high signal-to-noise (S/N) FUV spectra obtained by the {\it Hubble Space Telescope}/Cosmic Origins Spectrograph ({\it HST}/COS;  \citealt{Green2012}).

In this study, we investigate the properties of the warm-hot CGM traced by \ion{O}{6} and \ion{Ne}{8} in galaxies spanning a broad range in environments.
Section \ref{sec:data} summarizes the available data for both absorption spectroscopy and galaxy surveys, and presents the methodology to extract galaxy and absorption properties.
In Section \ref{sec:association}, we investigate the association between absorption systems and galaxies, and the spatial distribution and kinematic properties of the warm-hot CGM.
Further investigations of the dependence on galaxy properties are detailed in Section \ref{sec:gal_prop}. 
In particular, we divide the entire galaxy sample into sub-samples based on the stellar mass ($\mstar$) and star formation rate (SFR), and report different trends of absorption properties for different galaxies (Sections \ref{sec:radial2} and \ref{sec:radial_ne8}).
In Section \ref{sec:dis}, we compare the results with previous studies and discuss the implication for the origin of the warm-hot CGM.
The key findings are summarized in Section \ref{sec:summary}.
Throughout the paper, we adopt a $\Lambda$ cosmology to calculate physical distances, assuming $\Omega_{\rm m} = 0.3$, $\Omega_{\Lambda} = 0.7$, and a Hubble constant of $H_0 = 70\rm ~ km ~s^{-1}~ Mpc^{-1}$. 

\section{Data and Analysis}
\label{sec:data}
The CUBS program is designed to track the CGM evolution over a broad redshift range from $z\approx 0$ to $z\approx 1$ by combining high-quality QSO absorption spectra and deep galaxy survey data in the QSO fields \citep[see][for a complete description]{Chen:2020aa}.
Leveraging the CUBS data, we have established a new sample of 103 unique galaxies or galaxy groups at $0.4 \lesssim z\lesssim 0.7$, for which sensitive constraints can be obtained for both \ion{O}{6} and \ion{Ne}{8} absorption properties in their CGM. 
The three-tier CUBS galaxy redshift survey provides ultra-deep, deep-narrow, and shallow-wide coverage of the field around 15 NUV-bright QSOs at $z_{\rm QSO}>0.8$ (see \citealt{Chen:2020aa} for definitions). This enables a detailed characterization of intervening galaxies and their surrounding environments along the QSO sightlines \citep[e.g.,][]{Cooper:2021aa}.
The CUBS intervening galaxy sample, spanning a wide range in mass from $\log \mstar/\msun \approx 8$ to $\log\,\mstar/\msun\approx 11.5$ and a wide range in environments from isolated field galaxies to rich galaxy groups, provide a unique opportunity to study the \ion{O}{6} and \ion{Ne}{8}-bearing CGM in various galaxies. 

\subsection{QSO absorption spectra and galaxy survey data}
\label{sec:cubs_data}
For each CUBS QSO, high-S/N {\it HST}/COS spectra were obtained using the medium-resolution (FWHM $\approx 20 \kms$) G130M and G160M gratings and multiple central wavelengths to yield contiguous spectral coverage from 1100 \r{A} to 1800 \r{A} (PID$=$15163; PI: Chen).
The spectra are processed and coadded using custom software and the final combined spectra have typical S/N of $12\,-\,31$ per resolution element \citep[see details in][]{Chen:2018aa, Chen:2020aa}. 
The wavelength range of the final combined COS spectra provides simultaneous coverage of the \ion{O}{6} and \ion{Ne}{8} doublets over the redshift range $0.43 \lesssim z\lesssim 0.72$.

To establish a blind sample of intervening galaxies in this redshift range, 
we identify galaxies at projected distances $d\lesssim 1$ Mpc from the CUBS deep-narrow survey component that was carried out using a combination of the Multi Unit Spectroscopic Explorer (MUSE;\citealt{Bacon2010}), an integral field spectrograph on the Very Large Telescope (VLT) and Low Dispersion Survey Spectrograph 3 (LDSS3C), a high-efficiency optical multi-object spectrograph and imager on the Magellan Telescopes \citep[see][]{Chen:2020aa,Cooper:2021aa}. 
While the LDSS3C component reaches a limiting magnitude of $m_{r, {\rm AB}} \approx24$ out to $\approx 1$  Mpc at $z\approx 0.5$, the MUSE observations provide the deepest view in the inner 200 kpc around the QSO with a limiting pseudo-$r$ magnitude of $m_{r, \rm AB} \approx 26.0$. 
These spectroscopic data enable secure galaxy redshift measurements based on clearly detected spectral features (see \citealt{CUBSVI} for details; hereafter \citetalias{CUBSVI}).
We also include additional spectroscopic redshift survey data from the shallow-wide component, which reaches a limiting magnitude of $m_{r, {\rm AB}} \approx22.5$ out to $\approx 4$ Mpc at $z\approx 0.5$ using IMACS multislit observations.  These are useful for characterizing the large-scale environment of those galaxies found closer to the QSO sightlines. 

In addition to the spectroscopic survey data, deep optical have been obtained using LDSS3C and Inamori-Magellan Areal Camera \& Spectrograph (IMACS), a wide-field imager and spectrograph; \citealt{Dressler2011}), and near-infrared images using FourStar (a wide field near-infrared imaging camera; \citep{Persson:2013aa} on the Magellan telescopes.
These images reach 5-$\sigma$ limiting magnitudes of $\approx$ 26.0, 25.5, 25.5, 25.5, and 24.5 in the $g$-, $r$-, $i$-, $z$-, and $H$-band, respectively.
They provide additional constraints on the broad-band spectral energy distributions (SEDs) of individual galaxies.
In particular, we estimate the stellar mass for all galaxies with secure redshifts and five-band optical and near-infrared photometry available (DePalma et al. in prep.), using Bayesian Analysis of Galaxies for Physical Inference and Parameter EStimation package (BAGPIPES; \citealt{Carnall:2018aa}). 
We adopt the \citet{Kroupa:2001aa} initial mass function for this exercise (see \citetalias{CUBSVI} for detail).
This exercise leads to a sample of 1657 galaxies with $\log \mstar/\msun\approx 8-11.5$ at $z\approx 0.4-0.7$ in the CUBS galaxy survey.

The faintest galaxies in the CUBS galaxy sample are spectroscopically identified in the MUSE observations, the deepest spectroscopic survey component in the CUBS program.  As a result, the majority of these faint galaxies are not detected in the Magellan imaging survey, and only pseudo-$g$, $r$, and $i$ photometric measurements are available from MUSE \citep[e.g.,][]{Chen:2020aa}.  Constraints for $\mstar$ of these galaxies, therefore, rely on a color-magnitude analysis described in the following two paragraphs.  

To ensure that $\mstar$ is determined consistently for both bright and faint galaxies,
we define a calibration galaxy sample that has both pseudo-$g$, $r$, $i$ photometry from MUSE and $g,r,i,z,$ and $H$ photometric measurements from Magellan.
While using $g$-, $r$-, and $i$-band photometry alone does not provide as strong a constraint on $\mstar$ as using a combination of optical and near-infrared photometry, the three optical bands together can still place robust constraints on the rest-frame $B$-band absolute magnitudes ($M_B$) and $u-g$ colors for galaxies at $z\approx 0.4-0.7$.
The utility of this calibration sample is therefore to explore the possibility of constraining $\mstar$ using a combination of $M_B$ and $u-g$ for all galaxies in our sample \citep[e.g.,][]{Huang:2016aa}.

\begin{figure}
\begin{center}
\includegraphics[width=0.45\textwidth]{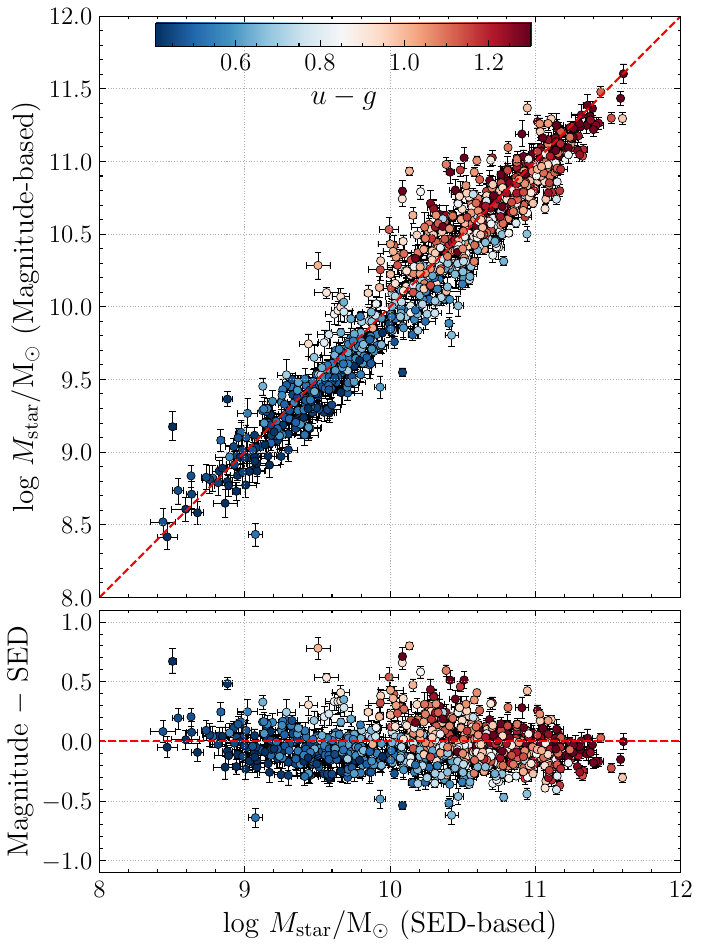}
\end{center}
\vskip -0.2in
\caption{Comparison between the SED-based and magnitude-based stellar masses.
The intrinsic scatter of the empirical relation between $\mstar$, $M_B$, and $u-g$ color is $\approx 0.2$ dex, which is determined using 910 galaxies at $z \approx 0.4-0.7$ with high-quality observed magnitudes, leading to derived $\mstar$ uncertainties of $< 0.1$ dex.}
\label{fig:mstar_relation}
\end{figure}

We first calculate the rest-frame $M_B$ and $u-g$ by incorporating only available observed $g$-, $r$-, and $i$-band photometry for each galaxy using the \textsc{kcorrect} software \citep{Blanton:2007aa}.
Next, we derive an empirical relation that best characterizes the correlation between BAGPIPES-inferred $\mstar$ and the rest-frame $M_B$ and $u-g$ color.
To account for the blue and red sequences in the galaxy population as a result of differences in star formation history \citep[e.g.,][]{Chen:2010aa,Johnson:2015aa}, we consider two different branches in the $\mstar$ and color-magnitude correlation.  We find that for blue galaxies with $u-g<0.9$
\begin{equation}
    \log \mstar/{\rm M_\odot} = -(0.36\pm0.02) M_{B} + (2.17\pm0.05)(u-g) \\
     +(1.1\pm0.3) ,
\end{equation}
and for red galaxies with $u-g\geq 0.9$
\begin{equation}    
   \log\mstar/\msun = -(0.36\pm0.02) M_{B} + (0.72\pm0.05)(u-g)+(2.4\pm0.3).
\end{equation}
The best-fit coefficients are determined using a Bayesian framework described in \citet{CUBSV}.
To determine the intrinsic scatter of this empirical relation, we select 910 galaxies at $z\approx0.4-0.7$ with well-measured magnitudes by filtering out galaxies with $\mstar$ uncertainties larger than 0.1 dex.
Using this high-quality galaxy sample, the residuals between the mean relation and the data exhibit a scatter of $\approx0.2$ dex over the covered stellar mass range of $\log \mstar/\msun \approx 8-11.5$ (Figure \ref{fig:mstar_relation}).  
The best-fit empirical relation is applied to all foreground galaxies for estimating $\mstar$ from the observed $g$, $r$, and $i$ magnitudes.

\begin{figure*}
\begin{center}
\includegraphics[width=0.98\textwidth]{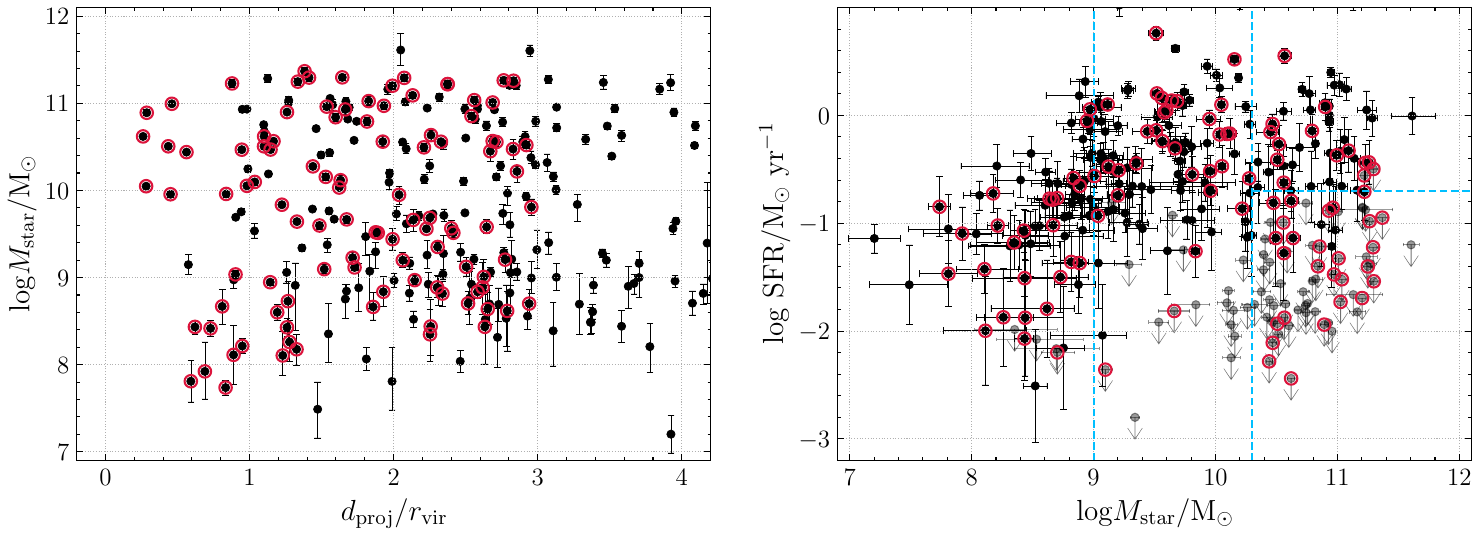}
\end{center}
\vskip -0.2in
\caption{Summary of the general galaxy characteristics of the CUBS sample at $0.4\lesssim z\lesssim 0.7$, including virial-normalized projected distance ($d_{\rm proj}/\rvir$), stellar mass ($M_{\rm star}$), and star formation rate (SFR).
Derived stellar masses have systematic uncertainties of $0.2$ dex over the covered $\mstar$.
For each galaxy group, the open red circle marks the galaxy with the smallest $\dproj/\rvir$ from the QSO sightline.
In this sample, most galaxies are star-forming with specific SFRs of $\geq10^{-11}\rm ~ yr^{-1}$.
In addition, the entire sample is split into four sub-samples in the following analysis, i.e., the dwarf, sub-$L^*$, massive star-forming, and massive quiescent galaxy samples (see Section \ref{sec:radial2}), which are classified using the dashed blue boundaries in the right panel.
}
\label{fig:sample}
\end{figure*}

In total, the CUBS galaxy sample consists of 11593 spectroscopically-identified foreground galaxies with available $g$, $r$, and $i$ magnitudes at $z_{\rm gal}>0.01$ and line-of-sight velocity of $>5000\,\kms$ away from the QSO emission redshift.
These galaxies are projected within $\approx 5$ Mpc around QSO sight lines.
For these galaxies, the derived $\mstar$ spans the range $\log \mstar/\msun \approx 8$ to 11.5.
We also compute an inferred dark matter halo mass using a stellar mass-halo mass (SMHM) relation at $z\lesssim 1$.  We adopt the relation from \citet{Behroozi:2019aa} and account for missing light described in  \citet{Kravtsov:2018aa} for all redshifts. 
The halo radius ($\rvir$) is approximated using $r_{200}$, within which the mean dark matter density is 200 times the cosmic critical density.

In addition to the stellar and halo mass estimation, we also extract SFRs using the detected nebular lines in galaxy spectra.
Specifically, we adopt the equivalent widths (EWs) of two nebular lines, the [\ion{O}{2}] doublet and H$\beta$, as tracers of SFR.
For these two features, we adopt the spectral windows defined in \citet{Yan:2006aa} to calculate the EWs.
For H$\beta$, we infer EW(H$\alpha$) using the anticipated line ratio of EW(H$\alpha$)/EW(H$\beta$)\,$\approx 2.8$.
Then, the line luminosities of [\ion{O}{2}] and H$\alpha$ are calculated by combining EW and rest-frame magnitudes ($u$ and $g$), and are converted into SFRs using the empirical conversion described in \citet{Kewley:2004aa} and \citet[][originated from \citealt{Kennicutt:1998aa}]{Fumagalli:2012aa} for [\ion{O}{2}] and H$\alpha$, respectively.
The [\ion{O}{2}]-based SFR is adopted for a galaxy when both [\ion{O}{2}] and H$\beta$ are available.
Among all 11593 galaxies in the CUBS program, 9130 and 2358 have SFRs determined based on [\ion{O}{2}] and H$\beta$, respectively.
For the remaining 105 galaxies, neither spectral feature is available due to gaps in the optical spectra.

\subsection{CUBS-midz: A new galaxy sample for probing the warm-hot CGM at $0.4\lesssim z\lesssim 0.7$}
\label{sec:galaxy_sample}

Here, we introduce a new sample of intermediate-redshift galaxies assembled from the entire CUBS galaxy sample for investigating the warm-hot CGM at $0.4\lesssim z\lesssim 0.7$ (hereafter designated as the CUBS-midz sample).  The redshift range is dictated by a simultaneous spectral coverage of the \ion{O}{6} and \ion{Ne}{8} doublets in the COS/FUV spectra.
A two-step procedure is adopted to select galaxies based on their proximity to the QSO sightlines.
First, we select all galaxies at projected distances ($\dproj$) less than $3\,r_{\rm vir}$ from the QSO sightlines, leading to a sample of 212 individual galaxies.
For these 212 galaxies, we then define overdense regions based on the number of galaxies with line-of-sight velocity separations of $|\Delta v| < 500~ \kms$ and projected separations of $d_{\rm proj}\lesssim 1$ Mpc from the QSO sightlines. 
We identify 60 isolated galaxies with no neighbors found in the immediate vicinity and 43 unique overdense regions with more than one galaxy identified in the search volume, leading to a total of 103 unique galaxies or galaxy groups in this new sample.
 
To further characterize the galaxy environment, we expand the search of associated galaxies beyond the initial $\dproj < 3\,r_{\rm vir}$ from the QSO sightlines. 
Specifically, we search for all nearby galaxies with projected separations of $\dproj\!\lesssim\!3\,r_{\rm vir}$ and velocity difference of $\lesssim 2\,v_{\rm vir}$ from the galaxies in the initial CUBS-midz sample.
This second iteration has led to the identification of additional neighbors, reducing the number of isolated galaxies to 50 and increasing the number of sight lines probing overdense environments to 53.
Hereafter, we use galaxy groups referring to overdense environments with multiple nearby galaxies, although some systems are ambiguous to be identified as galaxy groups or not.

In summary, this exercise establishes the final sample of 304 galaxies that form 103 unique galaxies or galaxy groups (summarized in Appendix Table \ref{tab:sample}).
The number of galaxies in the galaxy groups ranges from 2 to 21 with half of the groups containing more than four members.
Figure \ref{fig:sample} summarizes the ranges of $d_{\rm proj}/r_{\rm vir}$ to the QSO sightlines, \mstar, and SFR of the full CUBS-midz sample.
In particular, these galaxies span a range in $\log\mstar/\msun\approx 7$ to $11.5$ and SFR from $\lesssim 10^{-2}$ to $10 \rm ~M_\odot~yr^{-1}$.
For each system, the galaxy with the smallest $d_{\rm proj}/r_{\rm vir}$ to the QSO sightlines is highlighted with an open red circle.

In this CUBS-midz sample, several galaxies and galaxy groups have been studied in previous CUBS papers, but none have published constraints on the properties of \ion{Ne}{8}.
Specifically, the new sample includes the galaxy group hosting an ${\rm H}_2$-bearing damped \lya\ absorber (DLA) at $z_{\rm DLA}=0.57$  presented in CUBS II \citep{Boettcher:2021aa}, three galaxies or galaxy groups hosting a Lyman Limit System (LLS) at $z_{\rm LLS}=0.4-0.6$ presented in CUBS III \citep{Zahedy:2021aa}, and two galaxy groups each hosting a partial LLS (pLLS) at $z_{\rm pLLS}=0.47-0.54$ reported in CUBS IV \citep{Cooper:2021aa}.

\subsection{Absorption measurements}
\label{sec:abs}
Having established a new CUBS-midz galaxy sample at $z\!\approx \!0.4-0.7$, we proceed with searches for \ion{O}{6} and \ion{Ne}{8} absorption features within line of sight velocity of $1000 ~\kms$ for each galaxy or member galaxies in each galaxy group.
The velocity window corresponds to roughly twice the escape velocity of the most massive galaxies in the sample.
For our study, we search not only for \ion{O}{6} and \ion{Ne}{8} absorption features, but also for low-ionization transitions, such as \ion{H}{1}, \ion{O}{4}, and \ion{O}{5}, to guide the identifications of \ion{O}{6} and \ion{Ne}{8} based on matching kinematic absorption profiles.
While these low ions may not share similar component ratios with \ion{O}{6} or \ion{Ne}{8}, all detected \ion{O}{6} and \ion{Ne}{8} features in the new sample have detectable \ion{H}{1} or other low ionization transitions.
We require that the detected \ion{O}{6} or \ion{Ne}{8} absorption features exhibit matched doublet line ratios without significant contaminations, and that both lines are detected at $\gtrsim 2 \sigma$.
If one of the doublet members is significantly contaminated, then we consider this absorption as an upper limit.
If both members are contaminated, then no constraints are available for the absorption.

Next, we measure the absorption properties for \ion{O}{6} and \ion{Ne}{8} doublets using the Voigt profile fitting method described in \citet[][also see \citealt{Zahedy:2016aa}]{CUBSV}.
The observed features of \ion{O}{6} and \ion{Ne}{8} are decomposed into a minimum number of absorption components as required by the line profiles.
Then each component is modeled based on a Voigt function, characterized by the ion column density, Doppler $b$ parameter, and line centroid. A canonical 2:1 ratio is adopted to fit both doublet members simultaneously.
For non-detections, we obtain a 2-$\sigma$ upper limit for the column density with the line centroid fixed to either the systemic redshift of the galaxy or the redshift at the associated low ionization transitions.
For galaxy groups with non-detected absorption, the line centroid is fixed to the redshift of the galaxy with the smallest $\dproj/\rvir$.
The $b$ parameters of non-detections are fixed to 30 $\kms$, typical of known \ion{O}{6} and \ion{Ne}{8} absorbers from previous studies \citep[e.g.,][]{Werk:2013aa, Savage:2014aa, Johnson:2015aa, Zahedy:2019aa}.

We compute the line-of-sight properties of the warm-hot CGM based on the integrated absorption properties of \ion{O}{6} and \ion{Ne}{8}.  Specifically, we extract the total column density ($N$) as the summation of all individual absorption components.
Furthermore, we calculate the velocity centroid ($\varv_{\rm c}$), and the line-of-sight velocity dispersion ($\sigma_{\varv}$) for each ion following
\begin{eqnarray}
\begin{aligned}
\varv_{\rm c} =& \frac{\sum N(\varv)\varv}{N},\\
\sigma_{\varv}^2 =& \frac{\sum N(\varv)(\varv-\varv_c)^2}{N}, 
\end{aligned}
\end{eqnarray}
where $N(\varv)$ is the column density of the modeled Vogit profile in each velocity bin, and the zero velocity is selected to be the redshift of the galaxy with the smallest $\dvir$ (see discussion in Section \ref{sec:host}).

Note that the velocity dispersion calculated from the second moment has different physical implications in different absorption systems. 
When it is a single-component system, the calculated velocity dispersion is related to $b$ of that component according to $\sigma_\varv=b/\sqrt{2}$.
In this case, the measured velocity dispersion represents the combined thermal and non-thermal motions within an absorbing cloud.
If multiple components are detected, the velocity dispersion is dominated by the projected velocity difference between different components, which measures the large-scale relative motion between clumps.
In the following analysis, we designate absorption with $\sigma_\varv\geq 40~\kms$ as broad features, which mainly trace the inter-cloud kinematics, because the observed maximum $b$ value is $\approx 60 ~\kms$ for individual \ion{O}{6} components (e.g., \citealt{Werk:2013aa, Savage:2014aa, Zahedy:2019aa}).

\begin{figure}
\begin{center}
\includegraphics[width=0.45\textwidth]{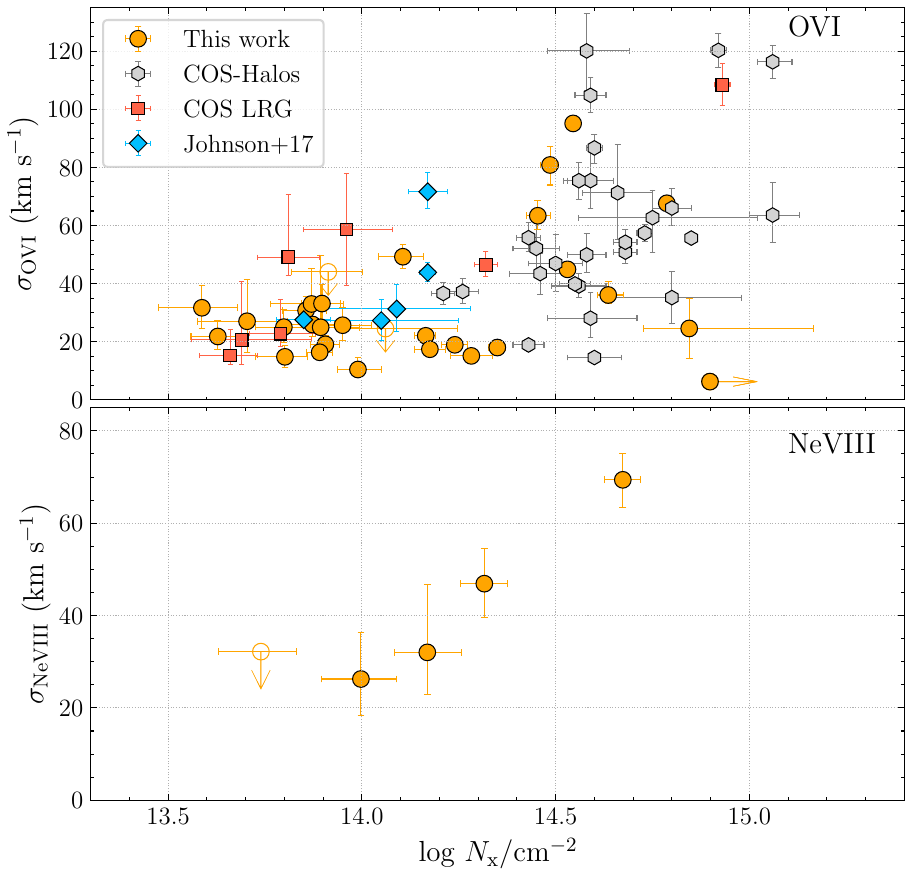}
\end{center}
\vskip -0.2in
\caption{Summary of the line-of-sight total column density and the velocity dispersion for detected \ion{O}{6} (top panel) and \ion{Ne}{8} features (bottom panel).
In this plot, we also include the COS-Halos (gray hexagons; \citealt{Werk:2013aa}), J17 (blue diamonds; \citealt{Johnson:2017aa}), and COS LRG samples \citep[red squares;][]{Zahedy:2019aa}.
All detected \ion{O}{6} features exhibit column densities of $\log N/\cmjj \gtrsim 13.5$, while broad features ($\sigma_{\varv}\geq40~\kms$) mostly have $\log N/\cmjj \gtrsim 14.5$.
For \ion{Ne}{8}, broad features have higher column densities, but the significance of this trend is limited by the small sample.
}
\label{fig:abs_sample}
\end{figure}

\section{The warm-hot CGM around galaxies}
\label{sec:association}
The analysis described in Section \ref{sec:data} returns detections of \ion{O}{6} in 30 galaxies or galaxy groups, while only five display associated \ion{Ne}{8} absorption features. 
For the remaining sample, we place a 2-$\sigma$ upper limit assuming $b=30\kms$ to the underlying ionic column densities at the locations of individual galaxies/galaxy groups. 
Figure \ref{fig:abs_sample} shows that the detected absorbers have ionic column densities spanning from $\log N/{\rm cm^{-2}}\!=\!13.5$ to 15.0 for both \ion{O}{6} and \ion{Ne}{8}, and the corresponding velocity dispersions vary from $\sigma_\varv\!\approx\!5$ to $120 \kms$.

Here, we also consider previously published galaxy samples for characterizing the \ion{O}{6}-bearing gas, including an IMACS survey from  \citet[][hereafter the J17 sample]{Johnson:2017aa}, the COS-Halos survey \citep{Werk:2013aa}, the COS-LRG survey \citep{Zahedy:2019aa}, and the CGM$^2$ survey \citep{Tchernyshyov:2022aa} for \ion{O}{6}.
The CASBaH sample is not included for comparisons in Figure \ref{fig:abs_sample} because neither the best-fit $b$ values or velocity dispersions of \ion{Ne}{8} were reported \citep{Burchett:2019aa}.  However, it will be included in comparisons of the column density profiles.
We find that broad absorption features with $\sigma_\varv\!>\!40\,\kms$ occur primarily in strong absorbers with $\log\,N/\cmjj\!>\!14.5$, indicating complicated kinematics in these systems.

In this section, we first explore whether and how any galaxy properties are correlated with the observed strength of the \ion{O}{6} and \ion{Ne}{8} absorbers. 
Specifically, we examine the significance of the correlation between absorption properties and a combination of different host galaxy assignments, including the galaxy at the smallest $d_{\rm proj}$, the galaxy at the smallest $r_{\rm vir}$-normalized $d_{\rm proj}$, and the most massive member in a galaxy group.
Next, we examine the covering fraction and gas kinematics of the warm-hot CGM probed by \ion{O}{6} and \ion{Ne}{8}.

\begin{figure}
\begin{center}
\includegraphics[width=0.46\textwidth]{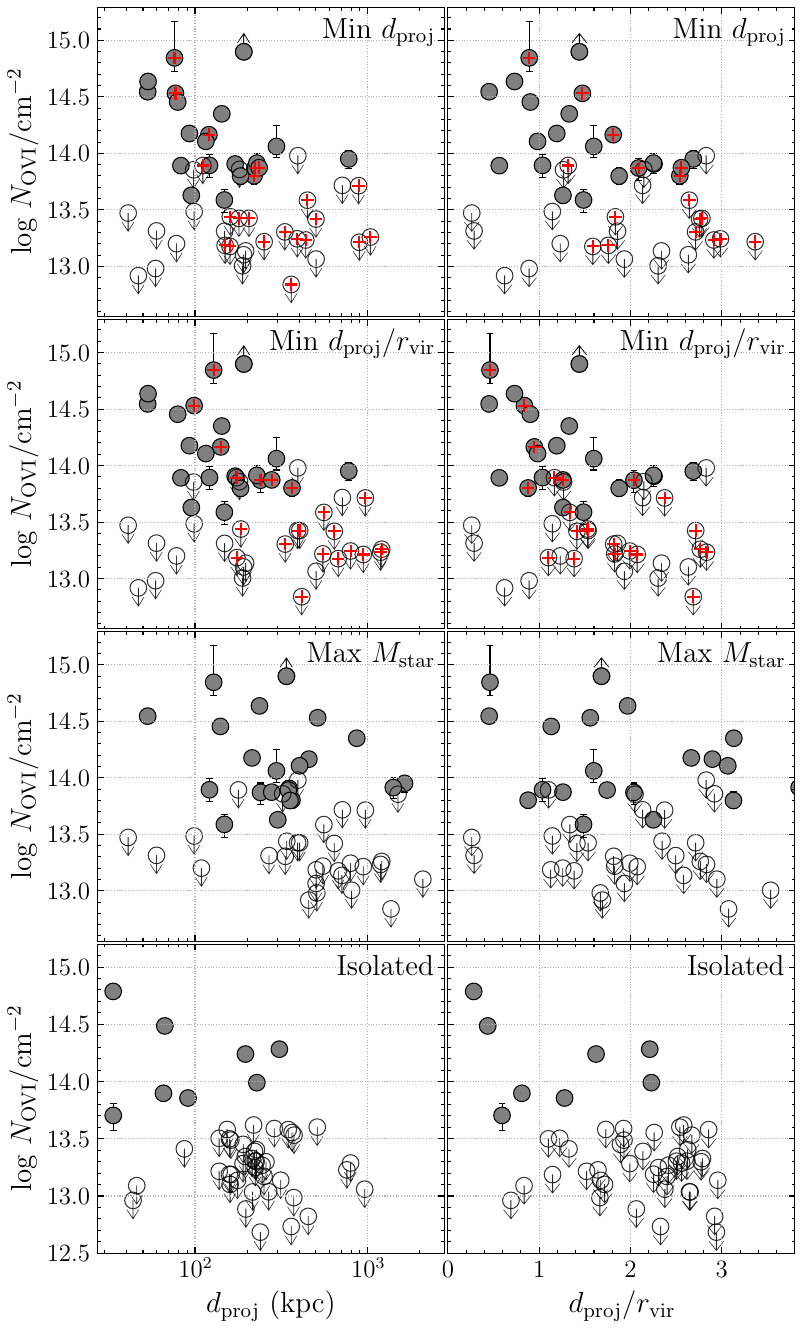}
\end{center}
\vskip -0.2in
\caption{Spatial profiles of \ion{O}{6} column density constructed based on different assignments of associated galaxies and projected absorber-galaxy distances. 
$N_{\rm OVI}$ is most strongly correlated with galaxies with the smallest $\dvir$, where $\rvir$ is calculated for individual galaxies (see Table \ref{tab:tau_comp} for statistics).
We present the dependence of column density on the projected distance and the $\rvir$-normalized distance in the left and right panels, respectively. 
Error bars represent 68\% confidence intervals of the column density measurements, and downward arrows represent the 95\% upper limits for non-detections.
In one case, the \ion{O}{6} absorber appears to be saturated, which is indicated by an upward arrow. 
In overdense galaxy environments, three different galaxy associations are explored, including the galaxy at the smallest $\dproj$, the galaxy at the smallest $\dproj/\rvir$, and the galaxy with the highest $\mstar$ from top to bottom in the first three rows.
The isolated galaxy sample is plotted in the lowest row for comparison.
In the first two rows, the physically closest galaxy is also the one with the smallest $\dproj/\rvir$ in 31 galaxy groups. 
For the remaining 22 groups, we highlight the difference in the associated galaxies using red plus symbols.
Overall, the combination of adopting the galaxies at the smallest $\dproj/\rvir$ and $\rvir$-normalized distance yields the tightest correlation (right panel in the second row).
}
\label{fig:gal_comp}
\end{figure}

\subsection{Connecting absorbers to galaxies in the presence of neighbors}
\label{sec:host}

As summarized in Section \ \ref{sec:galaxy_sample}, 53 unique galaxy groups containing between 2 and 21 members are identified in the CUBS-midz sample.  This sub-sample of galaxy groups provides an exciting opportunity to study the warm-hot CGM in overdense galaxy environments based on observations of \ion{O}{6} and \ion{Ne}{8} absorption.
In an overdense environment, however, the association of absorption features with individual members of the group becomes more ambiguous.
By comparing mean properties of the galaxy groups such as the center of mass, previous studies have shown that the \ion{O}{6}-bearing gas is more likely to be associated with individual galaxies rather than the intra-group medium \citep[e.g.,][]{Stocke:2014aa, Stocke:2017aa}.
However, other studies also suggest that the galactic environments could regulate the properties of absorbing gas \citep[e.g.,][]{Johnson:2015aa, Pointon:2017aa, Dutta:2021aa}. 
We will discuss the general impact of the galactic environment in future studies.
Here we apply the new CUBS-midz sample, including isolated and non-isolated galaxies to explore galaxy-absorber connections in the presence of neighbors.

\begin{table}[]
    \centering
    \caption{Summary of statistics in different galaxy associations}
    \begin{tabular}{llcc cc}
    \hline\hline
        Ion & Galaxy & \multicolumn{2}{c}{Physical$^a$} & \multicolumn{2}{c}{$\rvir$-Normalized$^a$} \\
        & & $\tau^b$ & Sig.$^b$ & $\tau^b$ & Sig.$^b$ \\
        \hline
        \ion{O}{6} & Min $\dproj$ & $-0.245$ & $2.0\sigma$ & $-0.277$ & $2.3\sigma$ \\
        \ion{O}{6} & Min $\dproj/\rvir$ & $-0.267$ & $2.2\sigma$ & $-0.298$ & $2.5\sigma$\\
        \ion{O}{6} & Max $\mstar$ & $-0.247$ & $1.9\sigma$ & $-0.036$ & $0.3\sigma$\\
        \ion{O}{6} & Isolated & $-0.186$ & $1.0\sigma$ & $-0.322$ & $1.8\sigma$\\
        \ion{Ne}{8} & Min $\dproj$ & $-0.088$ & $0.3\sigma$ & $-0.089$ & $0.3\sigma$\\
        \ion{Ne}{8} & Min $\dproj/\rvir$ & $-0.054$ & $0.2\sigma$ & $-0.085$ & $0.3\sigma$\\
        \ion{Ne}{8} & Max $\mstar$ & $-0.020$ & $0.1\sigma$ & $-0.062$ & $0.2\sigma$\\
        \ion{Ne}{8} & Isolated & $-0.200$ & $0.2\sigma$ & $-0.200$ & $0.2\sigma$\\
        \hline
    \end{tabular}
    \label{tab:tau_comp}
    \begin{flushleft}
$^a$ Dependence of $N_{\rm OVI}$ and $N_{\rm NeVIII}$ on $\dproj$ or $\dvir$.\\
$^b$ Generalized Kendall correlation coefficient $\tau$ and corresponding significance.
    \end{flushleft}
\end{table}

\begin{figure}
\begin{center}
\includegraphics[width=0.48\textwidth]{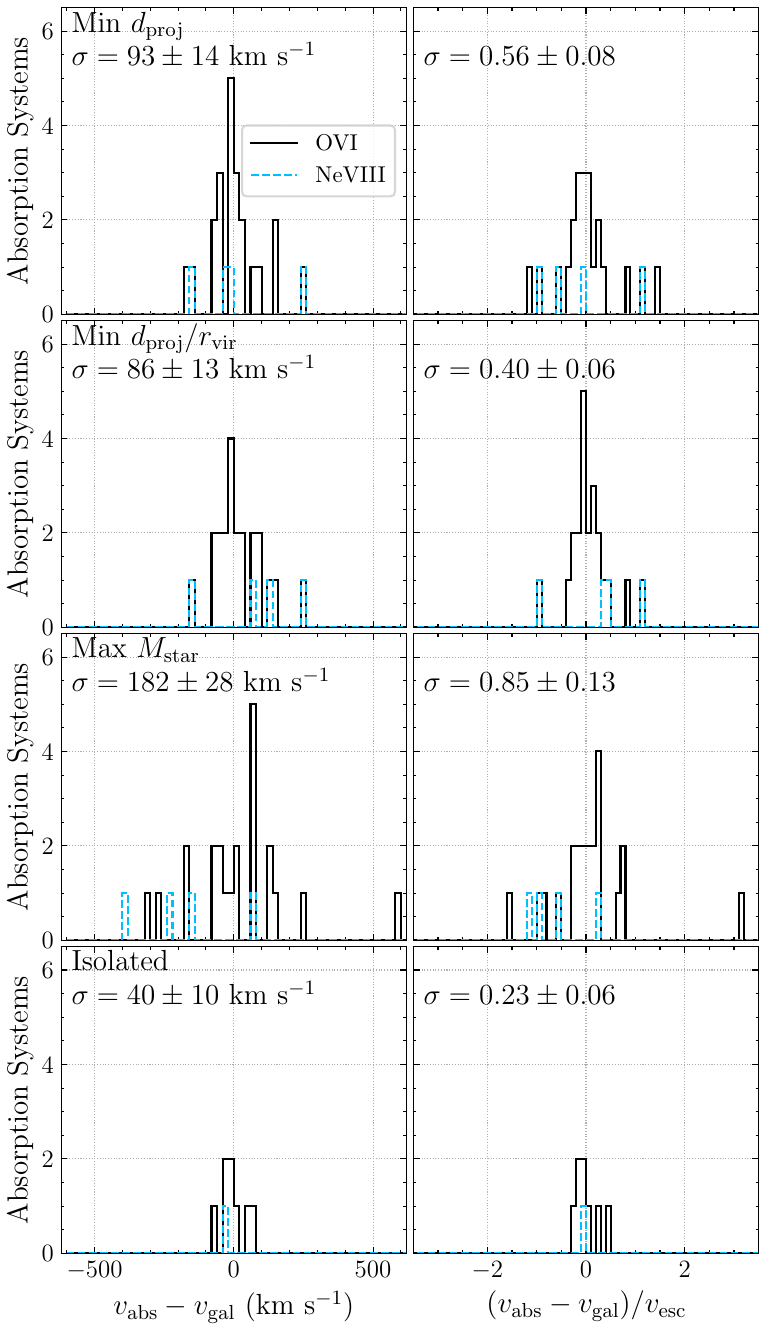}
\end{center}
\vskip -0.2in
\caption{Observed line-of-sight velocity distribution of \ion{O}{6} and \ion{Ne}{8} absorbers in the galaxy rest frame.  Solid histograms represent the \ion{O}{6} absorbers, while dashed histograms represent the \ion{Ne}{8} absorbers.  Galaxies are selected following the same criteria described in Figures \ref{fig:gal_comp} and \ref{fig:gal_comp_ne8}.
The right panels display escape velocity ($\varv_{\rm esc}$) normalized velocity dispersion.
The standard deviation ($\sigma$) of the \ion{O}{6} velocity distribution is marked in the top-right corner of each panel.
The narrower $\varv_{\rm esc}$-normalized velocity dispersion in the right panel in the second row suggests a tighter kinematic connection between the \ion{O}{6} absorbers and the galaxies at the smallest $\dproj/\rvir$ in a group environment.
Similar to the column density profiles displayed in Figure \ref{fig:gal_comp_ne8}, the small number of \ion{Ne}{8} absorbers detected in our search provides little distinction for their galaxy hosts. 
}
\label{fig:vel_sample}
\end{figure}

\begin{figure*}
\begin{center}
\includegraphics[width=0.48\textwidth]{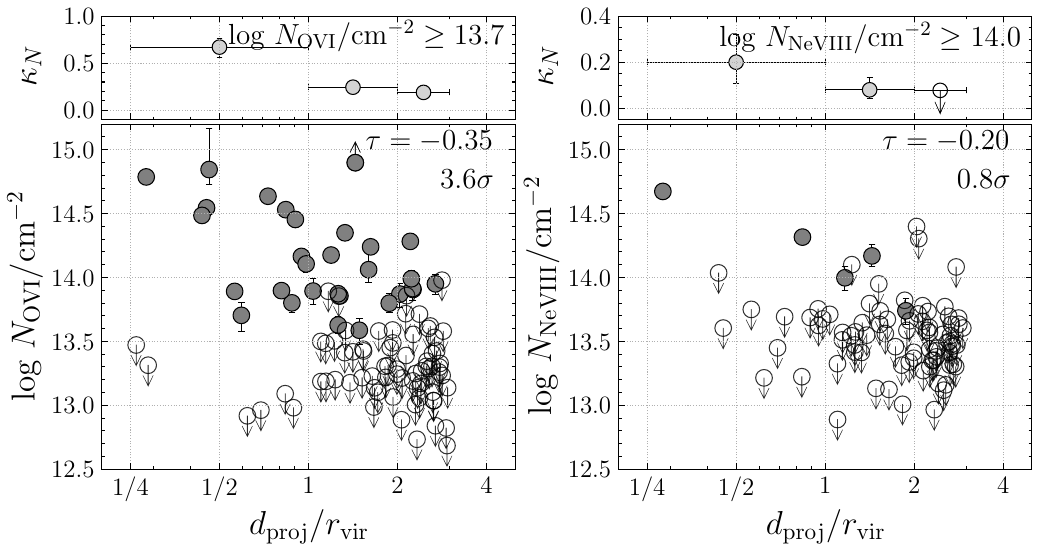}
\includegraphics[width=0.48\textwidth]{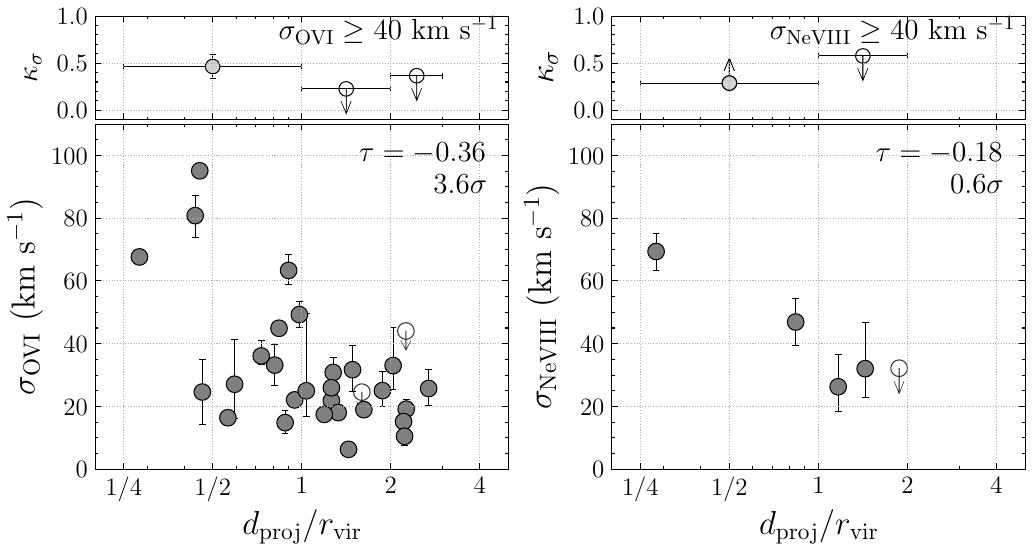}
\end{center}
\vskip -0.2in
\caption{Spatial profiles of column densities and velocity dispersion for \ion{O}{6} and \ion{Ne}{8}, together with the estimated covering fraction of high column density absorption systems with $\log N/\cmjj \geq 13.7$ and $14.0$ for \ion{O}{6} and \ion{Ne}{8}, respectively, at the top.  For velocity dispersions, the covering fractions are calculated for broad features with $\sigma_\varv \geq 40 \kms$.
}
\label{fig:N_radial}
\end{figure*}

In particular, we test the correlation between the light-of-sight absorption characterization and galaxy properties of different nearby galaxies.
In Figures \ref{fig:gal_comp}, we show respectively the spatial profiles of \ion{O}{6} absorption column densities for different host galaxy assignments.
Because few galaxies/galaxy groups display detectable \ion{Ne}{8} absorption, no clear trend can be established for any of the host galaxy assignments (see Figure \ref{fig:gal_comp_ne8} in the Appendix).
From the top to bottom panels, we consider the closest galaxy at the smallest $\dproj$ in each galaxy group, the galaxy with the closest $r_{\rm vir}$-normalized projected distance (minimum $\dproj/r_{\rm vir}$), and the most massive member of the group (maximum $\mstar$).
Here, virial radii are calculated for individual galaxies in each galaxy group.
As a comparison, we also show the isolated galaxy sample in the rightmost panels.
In the left and right panels, the radial profiles are plotted as functions of $\dproj$ and $\dproj/r_{\rm vir}$, respectively.
In each galaxy panel, we calculate the correlation coefficient $\tau$ using a generalized Kendall's rank order test including both measurements and limits, and quantify the significance of the correlation between the absorbing column density and the projected distance.
The results are summarized in Table \ref{tab:tau_comp}.

Among the three different galaxy assignments, linking the \ion{O}{6} absorption to the most massive member of the group exhibits the weakest correlation with either $d_{\rm proj}$ or $\dvir$. 
A generalized Kendall's test returns a rank coefficient of $|\tau|<0.25$ ($<2 \sigma$).
This weak correlation is similar to what was found for the cool CGM in \citepalias{CUBSVI}.
Furthermore, we examine whether galaxies at the smallest $\dproj$ or $\dproj/r_{\rm vir}$ exhibit a tighter correlation with the observed absorption properties.
In general, we find that adopting the galaxies at the smallest $\dproj/r_{\rm vir}$ slightly improves the significance of the anti-correlation, in comparison to adopting the galaxies at the smallest \dproj.
In addition, $N_{\rm OVI}$ correlates more strongly with $r_{\rm vir}$-normalized distance than with $\dproj$.

The finding of a strong correlation between the observed \ion{O}{6} absorbers and the galaxies at the smallest $\dproj/r_{\rm vir}$ is also supported by the bulk velocity distribution of the absorbers relative to the associated galaxies.
Figure \ref{fig:vel_sample} shows the velocity dispersions (left panels) and escape velocity ($\varv_{\rm esc}$) normalized velocity dispersion (right) between the observed \ion{O}{6} and \ion{Ne}{8} absorbers and the designated galaxies. 
We calculate $\varv_{\rm esc}$ based on the inferred dark halo mass for each galaxy in galaxy groups (see Section \,\ref{sec:data}), assuming the Navarro–Frenk–White (NFW) profile \citep{Navarro:1996aa}.
While no significant difference is seen in the velocity dispersions between \ion{O}{6} and galaxies at either the smallest \dproj\ or smallest  $\dproj/r_{\rm vir}$, the $\varv_{\rm esc}$-normalized velocity dispersions are notably narrower for galaxies at the smallest $\dproj/r_{\rm vir}$ (second panel in the right; showing a standard deviation of $0.40\pm0.06$) than those at the smallest \dproj\ (top-right panel; showing $0.56\pm0.08$). 
The difference is qualitatively consistent with the expectation that the observed absorbing gas is more closely connected to the galaxies at the smallest $\dproj/r_{\rm vir}$.
In the subsequent analysis, we adopt the galaxy at the smallest $r_{\rm vir}$-normalized $\dproj$ as the counterpart of the absorber to explore the connection between galaxies and absorption properties. 
Here, we ignore the potential impact on \ion{O}{6} or \ion{Ne}{8} due to nearby environments \citep[e.g.,][]{Dutta:2021aa}, which will be investigated in future CUBS works.

\subsection{Spatial properties of the warm-hot CGM probed by \ion{O}{6} and \ion{Ne}{8}}
\label{sec:radial}

By designating the galaxies at the smallest $\dproj/r_{\rm vir}$ as the primary driver of 
the observed line-of-sight absorption properties in a group environment, we proceed with a joint study of the spatial properties of the warm-hot CGM using the full sample of isolated galaxies and galaxy groups.
Using this full sample, we show the spatial profiles of $N_{\rm OVI}$ and $N_{\rm NeVIII}$ versus $\dproj/r_{\rm vir}$ in Figure \ref{fig:N_radial}.
On average, both $N_{\rm OVI}$ and $N_{\rm NeVIII}$ exhibit a general decline with increasing distance. 
Specifically, the detected \ion{O}{6} systems exhibit high column densities of $\log N/\cmjj \gtrsim 14.5$ at $<0.5\,\rvir$, together with two non-detections among six sightlines showing $2 \sigma$ upper limits of $\log N/\cmjj \lesssim 13.5$ within $\rvir$ (Figure \ref{fig:N_radial}).
In the outskirts at $\gtrsim 2 \rvir$, the overall strengths of detected \ion{O}{6} with column densities of $\log N/\cmjj \approx 13.5-14.0$ are sufficiently weak in comparison to the sensitivity of the data that a large fraction of sightlines show non-detections with a limiting column density of $\log N/\cmjj \lesssim 13.7$.

We determine the covering fractions of \ion{O}{6} and \ion{Ne}{8} absorbing gas above a column density threshold, $N_0$, versus projected distance.  These are presented at the top of each panel in Figure \ref{fig:N_radial}, which shows that the gas covering fraction also decreases with increasing distance.
For \ion{O}{6}, the adopted threshold for calculating the covering fraction is $\log N_0/\cmjj = 13.7$, which is larger than 95\% of the upper limits reported for the sample.  It allows us to include the majority of the sample for the covering fraction calculation.
The covering fraction is calculated assuming a binomial distribution applied to small samples \citep[see e.g.,][]{Gehrels:1986aa}.
Within $\rvir$, the covering fraction of \ion{O}{6} absorbers with $\log N_{\rm OVI}/\cmjj \geq 13.7$ is $\kappa_N =67\pm10\%$ (median and 1$\sigma$ uncertainty), and declines to $25\pm7\%$ and $19\pm6\%$ at $2\,\rvir$ and $3\,\rvir$, respectively.

For \ion{Ne}{8}, the column density profile resembles that of \ion{O}{6}, exhibiting higher column densities within $\rvir$.
However, significantly fewer detections are found and the data are not sufficiently sensitive for detecting \ion{Ne}{8} with $\log\,N/\cmjj\!\lesssim\!14$. 
Adopting a threshold of $\log N_0/\cmjj\!=\!14.0$, we estimate a covering fraction of $\kappa_N = 20_{-9}^{+12}\%$ for \ion{Ne}{8}-bearing gas within $\rvir$, which declines to $4_{-2}^{+3}\%$ at 1-3\,$\rvir$.
While the detected \ion{Ne}{8} sample is small, the decline in the covering fraction is significant.
To obtain a better constraint of the covering fraction, we combine the CUBS sample with literature samples, which will be presented in Section \ref{sec:cf}.

\begin{figure*}
\begin{center}
\includegraphics[width=0.98\textwidth]{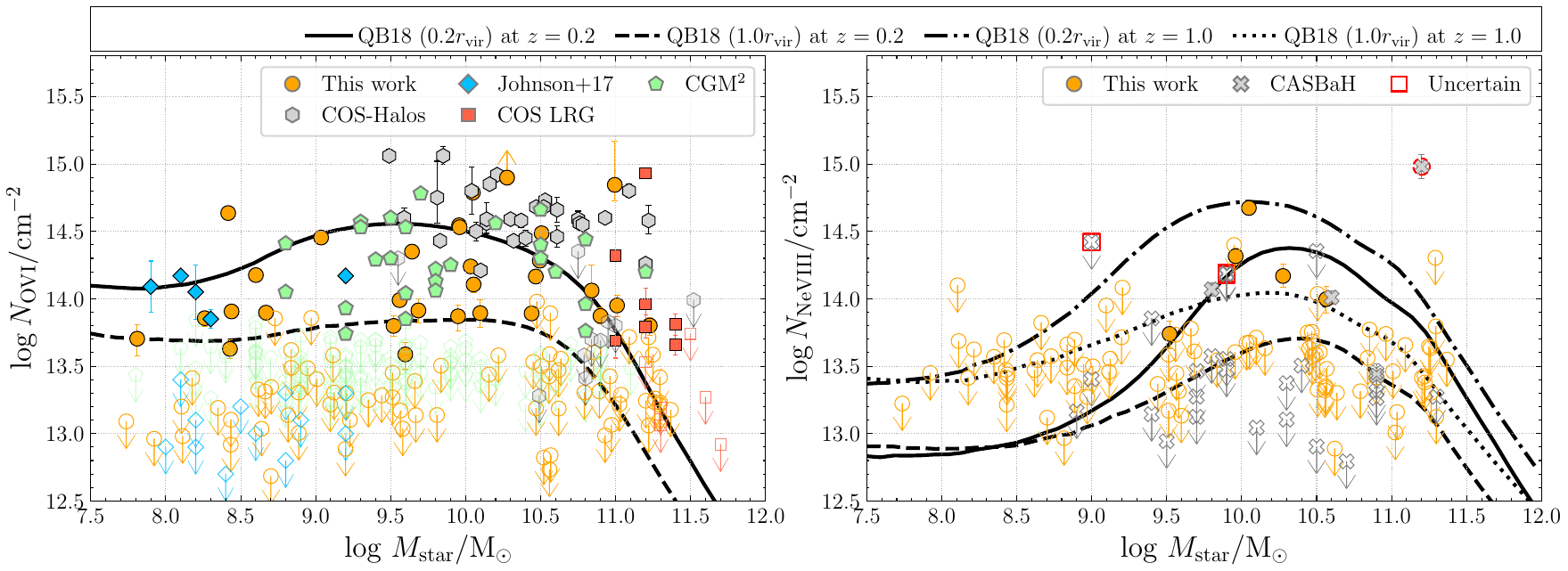}
\end{center}
\vskip -0.2in
\caption{Dependence of ion column density on \mstar\ for \ion{O}{6} (left panel) and \ion{Ne}{8} (right panel). 
Recall that in the presence of neighbors, we associate the observed line-of-sight absorption properties with the galaxy at the smallest $\dproj/\rvir$. 
Measurements from the literature, including COS-Halos \citep{Werk:2013aa}, J17 \citep{Johnson:2017aa}, COS-LRG \citep{Zahedy:2019aa}, CASBaH \citep{Burchett:2019aa}, and CGM$^2$ \citep{Tchernyshyov:2022aa} are also included.
In these samples, the maximum $\dvir$ are all smaller than 3.
Theoretical models adopted from \citetalias{Qu:2018aa} are plotted for comparisons (curves).
Both observations and theoretical models indicate that the observed ion abundances peak at $\sim 10^{10}\,\msun$ for both \ion{O}{6} and \ion{Ne}{8}.
For the CASBaH \ion{Ne}{8} sample, reported \ion{Ne}{8} systems with significant contaminations in \citet{Burchett:2019aa} are marked as uncertain systems in red squares (see text for details). 
One outlying strong \ion{Ne}{8} absorber (circled in red) near a massive galaxy of $\log\,\mstar/\msun=11.2$ containing an active galactic nucleus (AGN) is likely associated with AGN outflows \citep{Tripp:2011aa, Burchett:2019aa}.
}
\label{fig:N_ms}
\end{figure*}

As described in Section \ref{sec:abs}, the line-of-sight velocity dispersion $\sigma_\varv$ can serve as a tracer of the intra-halo kinematics, when the feature is broad $\geq 40\kms$.
Figure \ref{fig:N_radial} shows that there is a general decline of $\sigma_\varv$ from inner halos to the outskirts for both \ion{O}{6} and \ion{Ne}{8}.
In particular, broad \ion{Ne}{8} features are all associated with broad \ion{O}{6} features. 
All six broad \ion{O}{6} and two \ion{Ne}{8} absorbers are projected within $\rvir$, leading to a covering fraction of $\kappa_\sigma=46\pm13\%$ for these broad absorbers and $\kappa_\sigma > 29\%$ at the 2-$\sigma$ level of significance for \ion{O}{6} and \ion{Ne}{8}, respectively.
Beyond $\rvir$, we can only place a $2\sigma$ upper limit of $16\%$ and $58\%$ for broad \ion{O}{6} and \ion{Ne}{8}, respectively.
The difference in the covering fraction of broad \ion{O}{6} absorbers between within and beyond \rvir\ is statistically significant (at the level of $\approx 2\,\sigma$), suggesting that the gas kinematics is more complicated at $\dproj<\rvir$. 

\section{Dependence of the warm-hot CGM on galaxy properties}
\label{sec:gal_prop}

The new CUBS-midz sample includes isolated galaxies and galaxy groups containing 2 to 21 member galaxies. 
The analysis presented in Section\,\ref{sec:association} shows that in the presence of neighbors, the galaxy at the smallest $\dproj/\rvir$ is physically most connected to the absorbers identified along the QSO sightlines (see Section\,\ref{sec:association}).
In this section, we explore how the warm-hot CGM probed by \ion{O}{6} and \ion{Ne}{8} correlates with galaxy properties.
In particular, we focus on its relation to $\mstar$ and SFR, which characterize a galaxy's star formation history.

For literature samples, we adopt the reported total ion column densities when these measurements are available. When the best-fit Voigt profile parameters are reported for individual components, we compute a total column density by summing all components based on the Voigt profile fitting results.
For non-detections, we compute the corresponding 2-$\sigma$ limits based on the reported upper limits for all samples.
Finally, we calculate a velocity dispersion for each absorber in the literature samples based on the reported Voigt profile results, including the J17, COS-Halos, and COS-LRG samples. 
 For the literature samples, the adopted column densities and calculated velocity dispersions are also included in Figure \ref{fig:abs_sample} for comparisons.

Combining the CUBS-midz and literature samples leads to a joint sample of galaxies that span a range in \mstar\ from $\log \mstar/\msun \approx 8$ to $11.5$ over a redshift range from $z\approx 0.1$ to 1.0.
To ensure a consistent treatment of the absorber and galaxy association, we examine the public samples in search of possible neighbors and identify the galaxies at smallest $\dproj/\rvir$ as the designated galaxies associated with the reported \ion{O}{6} or \ion{Ne}{8} absorbers.
Virial radii are adopted from each sample directly for J17 and COS-LRG samples \citep{Johnson:2017aa, Zahedy:2019aa}, because the calculation methods are the same.
We recalculate the virial radius for the COS-Halos sample \citep{Werk:2013aa}.
However, because of inhomogeneous survey limits between the literature samples, discrepancies remain in characterizing the galaxy environments between the CUBS and these literature samples. The impact due to variations in survey depth will be discussed when combining these samples for further studies.

\begin{figure*}
\begin{center}
\includegraphics[width=0.98\textwidth]{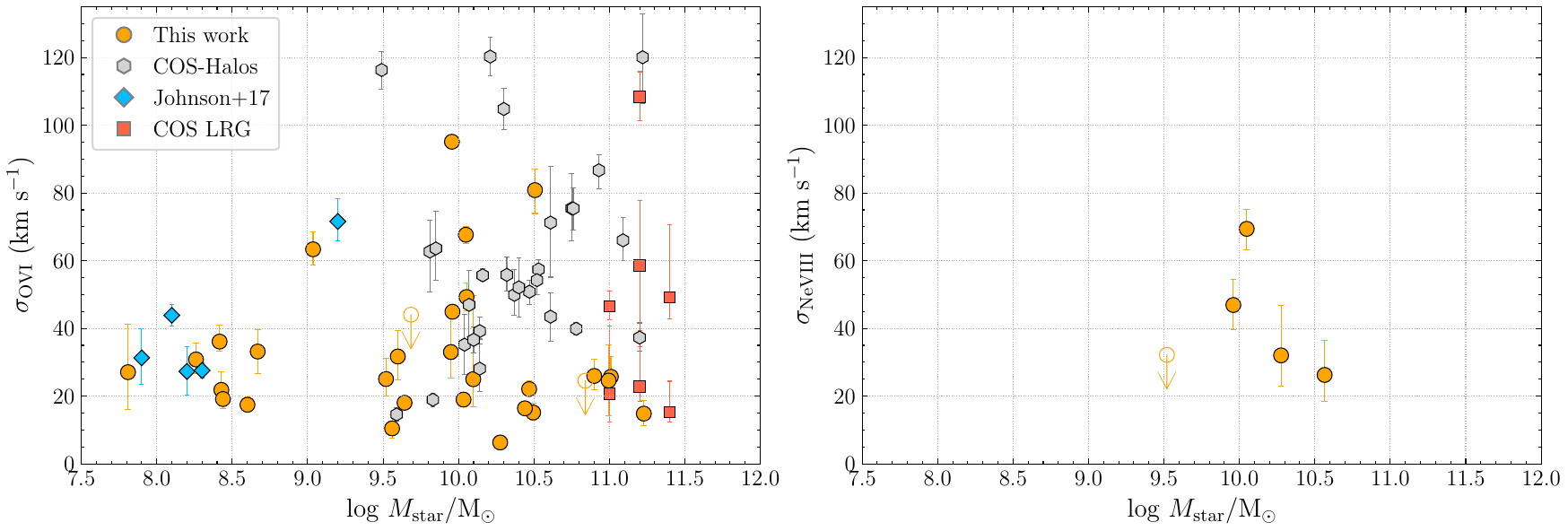}
\end{center}
\vskip -0.2in
\caption{The velocity dispersion dependence on the stellar mass of the galaxy with the minimal $\dproj/\rvir$ for \ion{O}{6} (left panel) and \ion{Ne}{8} (right panel).
Broad \ion{O}{6} absorption features of $\sigma_\varv \geq 40 \kms$ are detected at $\log \mstar/\msun \gtrsim 9.0$, while narrow features are found for all stellar masses.
}
\label{fig:b_ms}
\end{figure*}

\subsection{Dependence of \ion{O}{6} and \ion{Ne}{8}-bearing gas on the stellar mass}
\label{sec:ms}

In Figure \ref{fig:N_ms}, we examine the dependence of $N_{\rm OVI}$ and $N_{\rm NeVIII}$ on the stellar mass.
High $N_{\rm OVI}$ absorbers with $\log N/\cmjj \gtrsim 14.5$ are found surrounding galaxies of $\log \mstar/\msun \approx 9.5-10.5$, which is consistent with previous studies \citep[e.g.,][]{Werk:2014aa, Zahedy:2019aa, Tchernyshyov:2022aa}.
For a given $\mstar$, the observed $N_{\rm OVI}$ ranges from $2\sigma$ upper limits of $\log N/\cmjj\lesssim 13.0$ for non-detections to $\log N/\cmjj\gtrsim 14.0$ for measurements, which is driven by the radial decline shown in Figure \ref{fig:N_radial}.
For comparison, we plot the predicted $N_{\rm OVI}$ from semi-analytical CGM models for star-forming galaxies at $z=0.2$ and at $0.2 \rvir$ and $1.0 \rvir$ from \citet[][hereafter, \citetalias{Qu:2018aa}; also see Section \ref{sec:ratio}]{Qu:2018aa}. 
The predicted $N_{\rm OVI}$ for galaxies of different \mstar\ differs by a factor of $5-10$ between projected distances of $0.2~\rvir$ and $1.0~\rvir$.
The large scatter implies that the \ion{O}{6}-bearing gas may exhibit different radial distributions on different mass scales.  
We will explore this further in the following section.

For \ion{Ne}{8}, we include the CASBaH sample in the analysis.
However, we have to first construct a uniform sample by adopting the same identification criteria stated in Section \ref{sec:abs}.
For absorption features with significant contamination from interlopers, we consider
the reported $N_{\rm NeVIII}$ as upper limits.
This leaves four of the nine \ion{Ne}{8} absorbers reported by \citet{Burchett:2019aa} as detections and the remaining five as non-detections (red squares in Figure \ref{fig:N_ms}).
The four confirmed systems are PHL1377: 221\_15, PG1206+459: 178\_9, and PG1407+265: 245\_62, and FBQS0751+0919: 124\_25 (see Figure 2 of \citealt{Burchett:2019aa}).
In addition, we re-calculate the halo size for the CASBaH galaxies using their reported $\mstar$ and the method described in Section \ref{sec:cubs_data}.  This is necessary, because of the significant difference in the adopted SMHM relations (see details in \citealt{Wijers:2024aa}).

As shown in Figure \ref{fig:N_ms}, \ion{Ne}{8} features are primarily detected around galaxies with $\log \mstar / \msun \approx 9.5-11.0$.
We find that the strongest \ion{Ne}{8} features occur near galaxies of $\log \mstar/\msun\approx 10$, similar to \ion{O}{6}.
One outlying system with $\log N_{\rm NeVIII}/\cmjj=14.98\pm0.09$ in the CASBaH sample is likely associated with outflows from a massive post-starburst galaxy of $\log \mstar/\msun =11.2$ at $z=0.93$.  
This galaxy exhibits spectral features that are indicative of the presence of an AGN \citep{Tripp:2011aa}.
It is different from typical galaxies in the combined sample, but it is still included in the following analyses.

Next, we examine how the line-of-sight velocity dispersion, $\sigma_\varv$, depends on \mstar, which may shed light on the ionization mechanisms of the warm-hot CGM.
In PIE models, the typical density of \ion{O}{6} is $\approx 10^{-4} ~\cc$ \citep[e.g.,][]{Oppenheimer:2013aa, Stern:2018aa}, with an equilibrium temperature of $\log T/{\rm K} \approx4.5$ and a velocity dispersion of $\sigma_\varv \approx 4\kms$ due to thermal motions.  In contrast, under a CIE assumption, the ionization fraction of  \ion{O}{6} peaks at a temperature of $\log T/{\rm K} \approx 5.5$ with an anticipated thermal broadening of $\sigma_\varv \approx 12.8 \kms$.

In Figure \ref{fig:b_ms}, the velocity dispersion of \ion{O}{6} absorbers spans a wide range from $\approx 5$ to $120 \kms$, while \ion{Ne}{8} absorbers exhibit a minimum $\sigma_\varv\approx 20-30 \kms$.
We first note that the narrow \ion{O}{6} features with $\sigma_\varv \approx 5-10 \kms$ are likely photoionized or out of equilibrium \citep[e.g.,][]{Oppenheimer:2013aa}. 
On the other hand, the broad components with $\sigma_\varv \geq 40\kms$ are preferentially seen surrounding galaxies of $\log \mstar/\msun> 9$ for both \ion{O}{6} and \ion{Ne}{8}, and the majority of broad features are detected around galaxies of $\log \mstar/\msun\approx 10-11$.
For these massive galaxies, the observed gas velocity dispersion exhibits a large scatter similar to what is seen in the column density.  We attribute the observed large scatter to the complex gas kinematics at small $d_{\rm proj}/\rvir$ (Figure \ref{fig:N_radial}).

\begin{figure*}
\begin{center}
\includegraphics[width=0.98\textwidth]{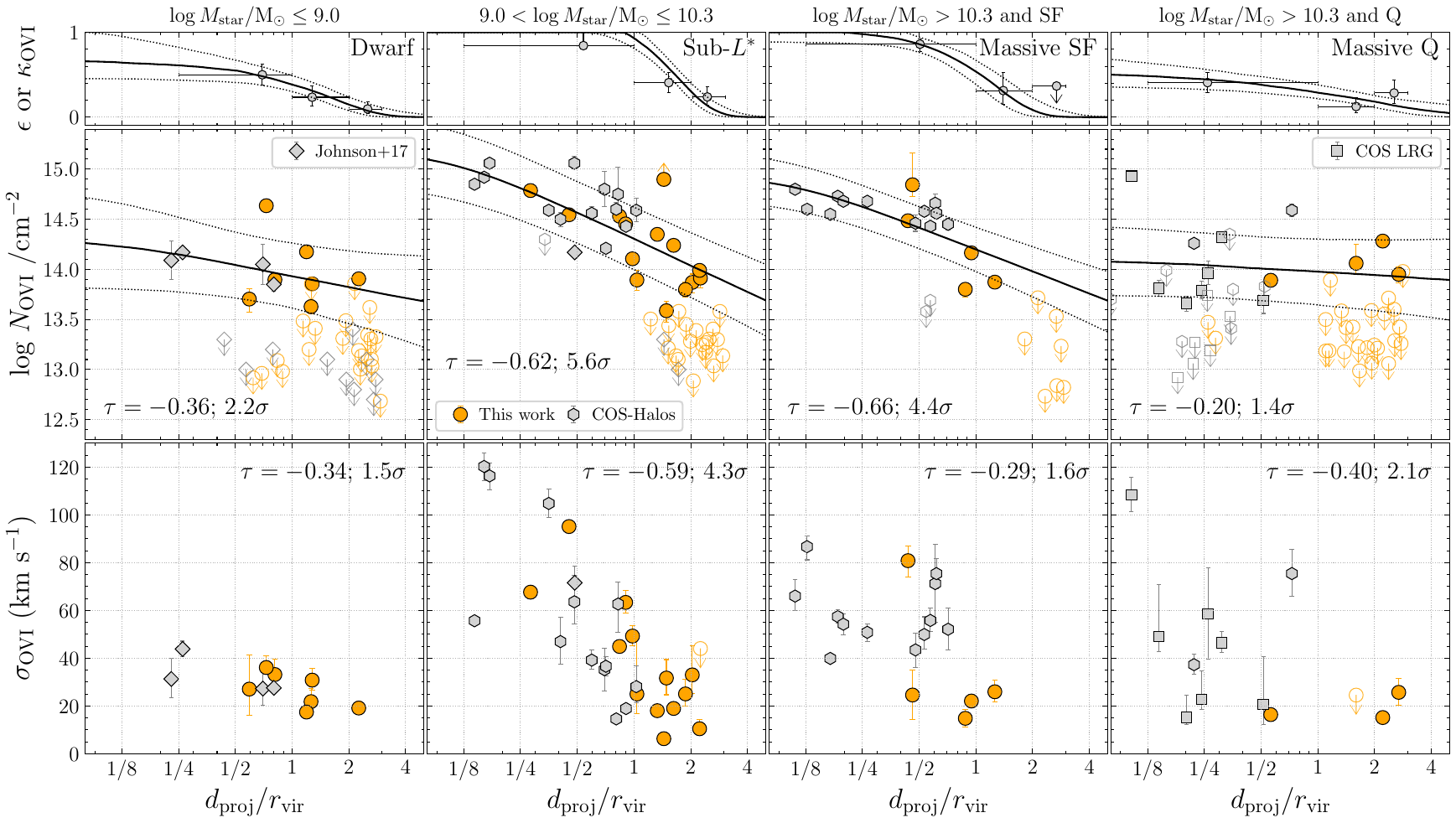}
\end{center}
\vskip -0.2in
\caption{Radial profiles of column density $N_{\rm OVI}$, covering fraction $\epsilon$, and velocity dispersion $\sigma_{\rm OVI}$ of \ion{O}{6}-bearing gas for low-mass dwarf, sub-$L^*$, massive star-forming, and massive quiescent galaxies from left to right columns, respectively.
In this plot, we include the literature samples in the analysis, the J17 dwarf sample (diamonds; \citealt{Johnson:2017aa}), COS-Halos (hexagons; \citealt{Werk:2013aa}), and COS LRG (squares; \citealt{Zahedy:2019aa}).
We calculate Kendall's rank-order coefficient $\tau$ to quantify the significance of the correlation, which is listed in each panel.
We fit the column density radial profile using a two-component model, including both the radial profiles of the column density (the middle row) and rate of incidence (the top row) to account for the clumpiness of the \ion{O}{6} gas (see the text for details).
The solid line in each panel shows the best-fit model, while the dotted lines represent the 1-$\sigma$ uncertainties.
Circles in the top panels represent the empirical covering fraction measurements obtained based on a threshold of $\log N/\cmjj \geq 13.7$.
The sub-$L^*$ and massive star-forming galaxy samples exhibit steeper declines with power law slopes of $\approx 0.9$, while the massive quiescent galaxy sample shows a flatter profile, consistent with a zero slope.
}
\label{fig:nb_rad_gals}
\end{figure*}

\subsection{Dependence of \ion{O}{6}-bearing gas on galaxy star formation history}
\label{sec:radial2}

Next, we examine whether the radial profiles of \ion{O}{6} column density and kinematics depend on a galaxy's star-formation history by dividing the CUBS galaxy sample into sub-samples in the $\mstar$ versus SFR parameter space (Figure \ref{fig:sample}). 
These include (i) a low-mass dwarf galaxy sample of 29 galaxies with $7.5\lesssim \log \mstar/\msun \leq 9.0$ and a median mass of $\langle\log \mstar/\msun\rangle =8.6$; (ii) a sub-$L_*$ sample of 34 galaxies with $9.0 <\log \mstar/\msun \leq 10.3$ and a median of $\langle\log \mstar/\msun\rangle =9.6$;
(iii) a massive, star-forming sample of 12 galaxies with $\log \mstar/\msun > 10.3$ and a median of $\langle\log \mstar/\msun\rangle = 10.7\pm0.3$; and (iv) a massive quiescent sample of 29 galaxies with $\log \mstar/\msun > 10.3$ and a median of $\langle\log \mstar/\msun\rangle = 10.9\pm0.3$.  Both the low-mass dwarf and the sub-$L_*$ sub-samples are predominantly star-forming galaxies with $\log {\rm sSFR/yr^{-1}} > -10.3$, while massive galaxies with ${\rm SFR}>0.2~\msun~\rm yr^{-1}$ are considered star-forming and quiescent otherwise.
This SFR threshold splits massive galaxies to have roughly equal numbers in each subsample together with literature samples.

Figure \ref{fig:nb_rad_gals} shows the radial profiles of column density and velocity dispersion for these sub-samples, i.e., dwarf, sub-$L^*$, massive star-forming, and massive quiescent galaxies from left to right respectively.
Here we also include the COS-Halos \citep{Werk:2013aa}, J17 \citep{Johnson:2017aa}, and COS-LRG samples \citep{Zahedy:2019aa} for improved statistics but leave out CGM$^2$ because the SFRs their galaxies and velocity dispersions of absorbers are not available \citep{Tchernyshyov:2022aa}.

The differences in the \ion{O}{6} column density profiles are significant between these four sub-samples.
Strong \ion{O}{6} absorbers with $\log N/\cmjj\gtrsim 14.5$ preferentially occur in the inner halo of sub-$L^*$ and massive star-forming galaxies, and the decline of $N_{\rm OVI}$ with increasing $d_{\rm proj}/\rvir$ is correspondingly significant in these two sub-samples.
This behavior explains that the on average higher \ion{O}{6} column density observed in the COS-Halos sample relative to the others may be understood as due to the QSO sightlines probing primarily 
the inner halo of galaxies with $\log \mstar/\msun\approx 10$.

We adopt the generalized Kendall correlation coefficient ($\tau$; Figure \ref{fig:nb_rad_gals}) to quantify the significance of the anti-correlation in the presence of upper limits and find a significance level of $> 4\sigma$ for the sub-$L^*$ and massive star-forming samples.
In contrast, the dwarf and massive quiescent sub-samples exhibit weaker correlations between $N_{\rm OVI}$ and $\dvir$.

To investigate this difference quantitatively, we perform an analysis under a Bayesian framework to obtain a best-fit radial profile of both ion column density and covering fraction for each sub-sample.
This two-component model is developed to account for the possible clumpy nature of the \ion{O}{6}-bearing gas (also see \citealt{Huang:2021aa}), and 
motivated by the observation that a single continuous probability distribution of $N_{\rm OVI}$ cannot simultaneously reproduce the observed high-$N_{\rm OVI}$ systems and non-detections within $\rvir$ (Section \ref{sec:radial} and Figure \ref{fig:N_radial}).
Instead, the difference between detections and non-detections can be explained by a non-unity covering fraction, $\epsilon$.  We note that $\epsilon$ is a theoretical parameter that depicts the presence or absence of \ion{O}{6} gas, while $\kappa$ is an empirical number that is affected by the sensitivity of the data and therefore is only meaningful with an associated column density threshold. 
In practice, the modeling of $\epsilon$ may be affected by the detection limits, when QSO spectra have low S/N ratios.
Here, we note the spectral S/N is sufficient to characterize $\epsilon$ for \ion{O}{6} in the CUBS program.

The two-component model is implemented by combining the spatial variations of $\epsilon$ and $N_{\rm OVI}$ as a function of $\dvir$.
We adopt a power-law radial profile for the expected \ion{O}{6} column density $\hat{N}$ 
\begin{eqnarray}
\hat{N}(x) = \hat{N}_0 x^{-\alpha},
\end{eqnarray}
where $x=\dvir$, $\hat{N}_0$ is the expected \ion{O}{6} column density at $\rvir$ and $\alpha$ is the slope.
The power law is selected because detected column densities do not exhibit a sharp decline, and it is also typically adopted in such modeling \citep[e.g.,][]{Chen:1998aa, Huang:2021aa}.
Here, we only fit data with $x$ in the range of $0.1$ to $3$, so we can ignore the singularity at $x=0$ for the power law.

For the covering fraction, we assume a modified exponential decline of 
\begin{eqnarray}
\epsilon (x) = {\rm Min} \left[1,  \epsilon_0 \exp\left(-\left(\frac{x}{x_0}\right) ^\beta\right)\right],
\end{eqnarray}
where $x_0$ is the $\rvir$-normalized scale radius of the exponential model, $\beta$ is the free index regulating the speed of decline, and $\epsilon_0$ is the covering fraction at the galaxy center.  Because the mathematical form as given allows $\epsilon_0$ to be larger than unity, we impose a ceiling at $\epsilon=1$. 
This modified exponential function is motivated by the sharp decline in the covering fraction (Figure \ref{fig:N_radial}), which has also been seen for \ion{Mg}{2}-bearing gas \citep{Huang:2021aa, Schroetter:2021aa}. 
In our model, if $\beta > 1$, then $\epsilon$ would decline faster than an exponential function.  Therefore, this modified exponential function provides the flexibility to capture different degrees of declining rate.

The likelihood of producing an observed data set $D$ that contains $n$ measurements, $m$ non-detections, and $l$ saturated lines under the two-population model $M$ consisting of parameters $\hat{N}_0$, $\alpha$, $\epsilon_0$, $x_0$, and $\beta$ is the joint product of (i) the probability of measuring $N_i$ for an expected $\hat{N}$ and (ii) the probability of the model values occurring within the range of allowed upper or lower limits by the data.  We construct a likelihood function following this joint probability,

\begin{eqnarray}
    & \mathcal{L}(D|M) \propto  \prod_{i=1}^{n} \frac{\epsilon}{\sigma \sqrt{2\pi}} \exp\left(-\frac{(\log N_i-\log \hat{N})^2}{2\sigma^2}\right) \notag \\ 
    & \times   \prod_{i=1}^{m} \left[\left(\frac{\epsilon}{\sigma_{\rm p} \sqrt{2\pi}}\int_{-\infty}^{\log N_i^u}  \exp\left(-\frac{(y-\log \hat{N})^2}{2\sigma_{\rm p}^2}\right)  {\rm d} y\right) 
    + (1 - \epsilon)\right] \notag \\
    & \times  \prod_{i=1}^{l} \frac{\epsilon}{\sigma_{\rm p}  \sqrt{2\pi}}\int^{+\infty}_{\log N_i^l}  \exp\left(-\frac{(y-\log \hat{N})^2}{2\sigma_{\rm p} ^2}\right)  {\rm d} y,
\end{eqnarray}
where $y=\log N/\cmjj$, $\sigma^2=\sigma_i^2+\sigma_{\rm p}^2$ with $\sigma_i$ representing the measurement uncertainty and $\sigma_{\rm p}$ representing the intrinsic scatter. 
The symbols $N_i$, $N_i^u$, and $N_i^l$ represent measurements, 95\% upper limits in case of non-detections, and 95\% lower limits in case of saturated lines, respectively.
The term ($1-\epsilon$) associated with upper limits accounts for the probability that the non-detection is from the region without \ion{O}{6}-bearing gas.

We assume a uniform prior and construct the posteriors of individual model parameters $\log \hat{N}_0$, $\alpha$, $\epsilon_0$, $x_0$, $\beta$, and $\sigma_{\rm p}$ based on Markov chain Monte Carlo (MCMC), which is implemented using the \texttt{emcee} package \citep{Foreman-Mackey:2013aa}.
The best-fit parameters are summarized in Table \ref{tab:fits}.
The median and 68\% interval of the best-fit model are presented as solid and dashed lines in Figure \ref{fig:nb_rad_gals}.

\begin{table*}
    \centering
    \caption{The best-fit models of the radial profiles of the column density for different galaxy samples.}
    \label{tab:fits}
    \setlength{\tabcolsep}{14pt}
    \begin{tabular}{lccccccccc}
    \hline\hline
         Sample & $\log N_0/\cmjj$ & $\alpha$ & ${\epsilon_0}$ & $x_0$\, (\rvir) & $\beta$ & $\sigma_{\rm p}$ \\
         \hline
         Dwarf & $13.93_{-0.13}^{+0.12}$ & $0.39_{-0.39}^{+0.44}$ & $0.63_{-0.21}^{+0.46}$ & $1.5_{-0.7}^{+0.5}$ & $1.7_{-0.9}^{+1.4}$ & $0.33_{-0.07}^{+0.10}$ \\
         Sub-$L^*$ & $14.32\pm0.07$ & $0.86\pm0.19$ & $[0.95, 2.0]^a$ & $1.6\pm0.3$ & $2.1_{-0.7}^{+0.9}$ & $0.32_{-0.05}^{+0.06}$ \\
         Massive SF & $14.20\pm0.09$ & $0.74\pm0.21$ & $1.04_{-0.17}^{+0.43}$ & $1.2_{-0.3}^{+0.4}$ & $2.0_{-0.9}^{+1.2}$ & $0.21_{-0.04}^{+0.06}$\\
         Massive Q & $13.97_{-0.12}^{+0.13}$ & $0.12_{-0.25}^{+0.23}$ & $0.60_{-0.20}^{+0.35}$ & $[0.1, 3.0]^b$ & $0.8_{-0.4}^{+1.0}$ & $0.3$ (fixed)$^c$ \\
         \hline
    \end{tabular}
    \begin{flushleft}
    $^a$ For sub-$L^*$ galaxies, the data do not provide distinguishing power a $\epsilon_0 > 1$, because of a high covering fraction of \ion{O}{6}-bearing gas in the inner halo. The upper bound of 2.0 is the fixed boundary for $\epsilon_0$. \\
    $^b$ For massive quiescent galaxies, a roughly flat covering fraction cannot distinguish $x_0$ within the allowed range of [0.1, 3.0]. \\
    $^c$ In the massive quiescent sample, $\sigma_{\rm p}> 0.3$ dex cannot be distinguished by the data, so we fixed $\sigma_{\rm p} =  0.3$ dex, similar to other sub-samples.
    \end{flushleft}
\end{table*}

\begin{figure*}
\begin{center}
\includegraphics[width=0.98\textwidth]{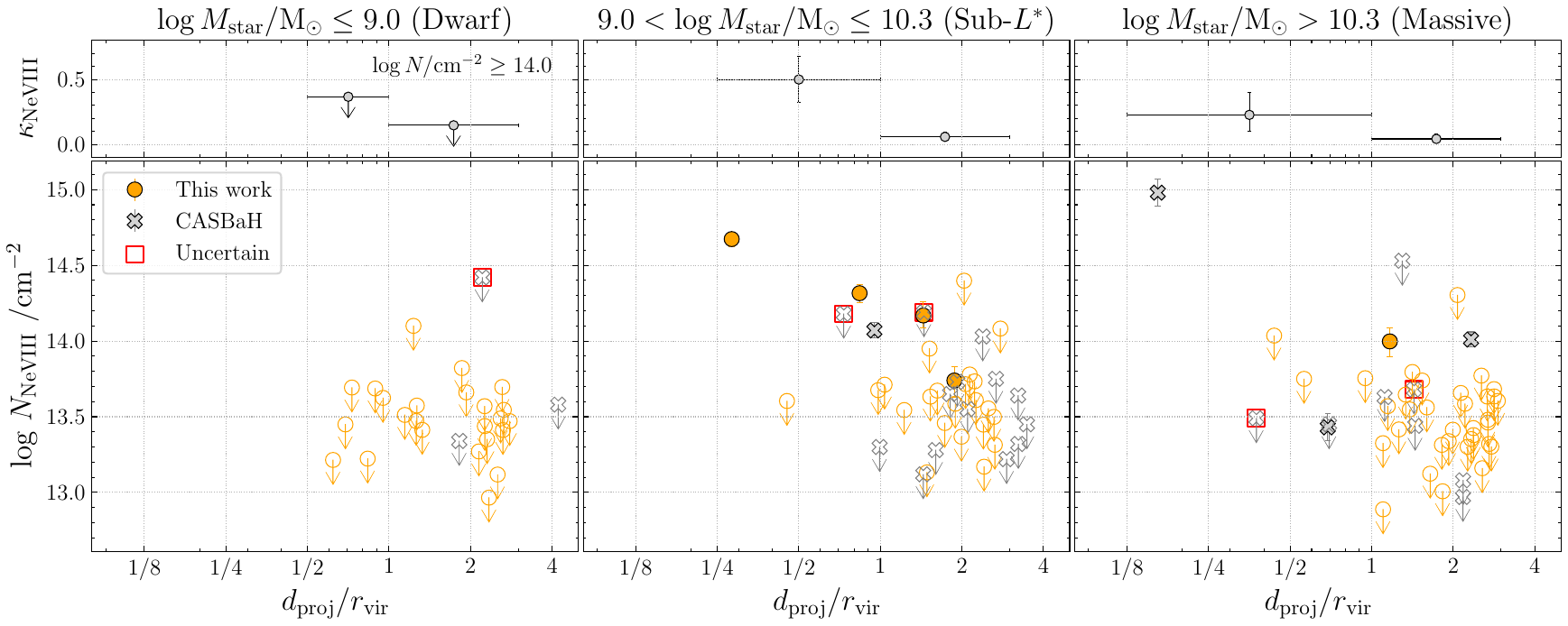}
\end{center}
\vskip -0.2in
\caption{The radial dependence of \ion{Ne}{8} column density and covering fraction for dwarf, sub-$L^*$, and massive galaxies.
The CASBaH sample is also plotted (crosses) with uncertain/contaminated detections marked with a red square (see details in Section \ref{sec:ms}).
The covering fractions are calculated for a threshold of $\log N/\cmjj \geq 14.0$.
High column density \ion{Ne}{8} systems are more concentrated in sub-$L^*$ galaxy halos.
}
\label{fig:radial_ne8}
\end{figure*}

Figure \ref{fig:nb_rad_gals} and Table \ref{tab:fits} show that while the sub-$L^*$ and massive star-forming sub-samples share a similar power law slope for the \ion{O}{6}-bearing gas ($0.86\pm0.19$ versus $0.75_{-0.18}^{+0.19}$), the massive quiescent sub-sample exhibit a flatter slope of $0.10_{-0.22}^{+0.24}$, respectively.
The dwarf galaxy sample shows a flatter slope of $0.39_{-0.39}^{+0.44}$, but its difference from sub-$L^*$ and massive star-forming galaxies is limited by the small sample size. 
Within a similar mass range, star-forming galaxies exhibit a steeper slope than the quiescent ones at a significance level of $\approx 1.6\,\sigma$.
This implies a fundamental difference in the spatial distribution of \ion{O}{6} between star-forming and quiescent galaxies. 

We also compare the observed covering fraction at a threshold of $\log N_{\rm OVI}/\cmjj = 13.7$, $\kappa$, with the model-predicted covering fraction, $\epsilon$, in Figure \ref{fig:nb_rad_gals}.
Similar to the column density profile, sub-$L^*$ and massive star-forming galaxies exhibit covering fractions consistent with unity at $\lesssim 0.5\,\rvir$ and declining to $\lesssim 20\%$ beyond $2\,\rvir$.
At the same time, dwarf galaxies exhibit a maximum covering fraction of $\approx 60\%$ at $0.5-1.0\rvir$.
The decline of covering fractions is significant for the three star-forming galaxy samples, with $\beta\approx2$.
The scale radius ($r_0$) is about $1.5\,\rvir$, suggesting an enhancement of the \ion{O}{6}-bearing gas within $\approx \rvir$ due to the nearby star-forming galaxies.

In contrast, the massive quiescent galaxy sample exhibits a roughly constant covering fraction of $\approx 30\%$ within $3\,\rvir$.
The best-fit $\beta= 0.7_{-0.4}^{+1.2}$ is smaller than the three star-forming samples $\approx 2.0$, suggesting a shallower decline.
In this case, the scale radius $r_0$ is unconstrained, because of the roughly constant covering fraction.

The difference between star-forming and quiescent galaxy CGM has also been seen in low ions.
For example, the rest-frame absorption equivalent width (EW) of \ion{Mg}{2} shows a power-law slope of $1.03\pm0.22$ for the star-forming sample, while the passive galaxy sample exhibits a slope of $0.12\pm0.24$ \citep[][]{Huang:2021aa}.
The cool and warm-hot gas in the CGM both exhibit concentrations in the inner halo of star-forming galaxies, compared to passive galaxies.
However, the covering fraction of $\approx 50\%$ for high column density absorbers occurs at $1-1.5 \rvir$ for \ion{O}{6}, while this radius is $0.5 \rvir$ for \ion{Mg}{2}, indicating changing ionization states of the gas toward halo outskirts.

Figure \ref{fig:nb_rad_gals} also shows a larger velocity dispersion within $\rvir$ than in the outskirts for all four sub-samples.
Here, we adopt the generalized Kendall's $\tau$ test to quantify the degree of anti-correlations for these samples.
The sub-$L^*$ sample exhibits a significant decline ($4.3\sigma$) from the maximum of $120 \kms$ within $0.2\,\rvir$ to $\approx 20 \kms$ at $1-2\,\rvir$, with all broad features ($\geq40\kms$) projected within the virial radius.
Anti-correlations are observed for the other three sub-samples, but less significant $\lesssim 2 \sigma$.
In addition, we notice almost all \ion{O}{6} features are broad within $0.5 \rvir$ for sub-$L^*$ and massive star-forming galaxies, while there are also narrow features within $0.5\,\rvir$ for massive quiescent galaxies.
The joint observations of enhanced $N_{\rm OVI}$ and broader $\sigma_\varv$ at $\lesssim 0.5\,\rvir$ around star-forming galaxies support that these high-$N_{\rm OVI}$ at small $\dvir$ absorbers do indeed originate in the galaxy halo, rather than appearing by projection \citep[cf.][]{Ho:2021aa}.

\subsection{Dependence of \ion{Ne}{8}-bearing gas on galaxy properties}
\label{sec:radial_ne8}
Similar to \ion{O}{6}, we consider the dependence of \ion{Ne}{8} profiles on galaxy properties.
For \ion{Ne}{8}, limited by the small number of detections, we only divided the uniform combined sample (Section \ref{sec:ms}) into three sub-samples based on $\mstar$, separated at $\log \mstar/\msun = 9.0$ and $10.3$.
In Figure \ref{fig:radial_ne8}, we present radial profiles of $N_{\rm NeVIII}$ as a function of $\dvir$ for these three sub-samples.

\begin{figure*}[t]
\begin{center}
\includegraphics[width=0.98\textwidth]{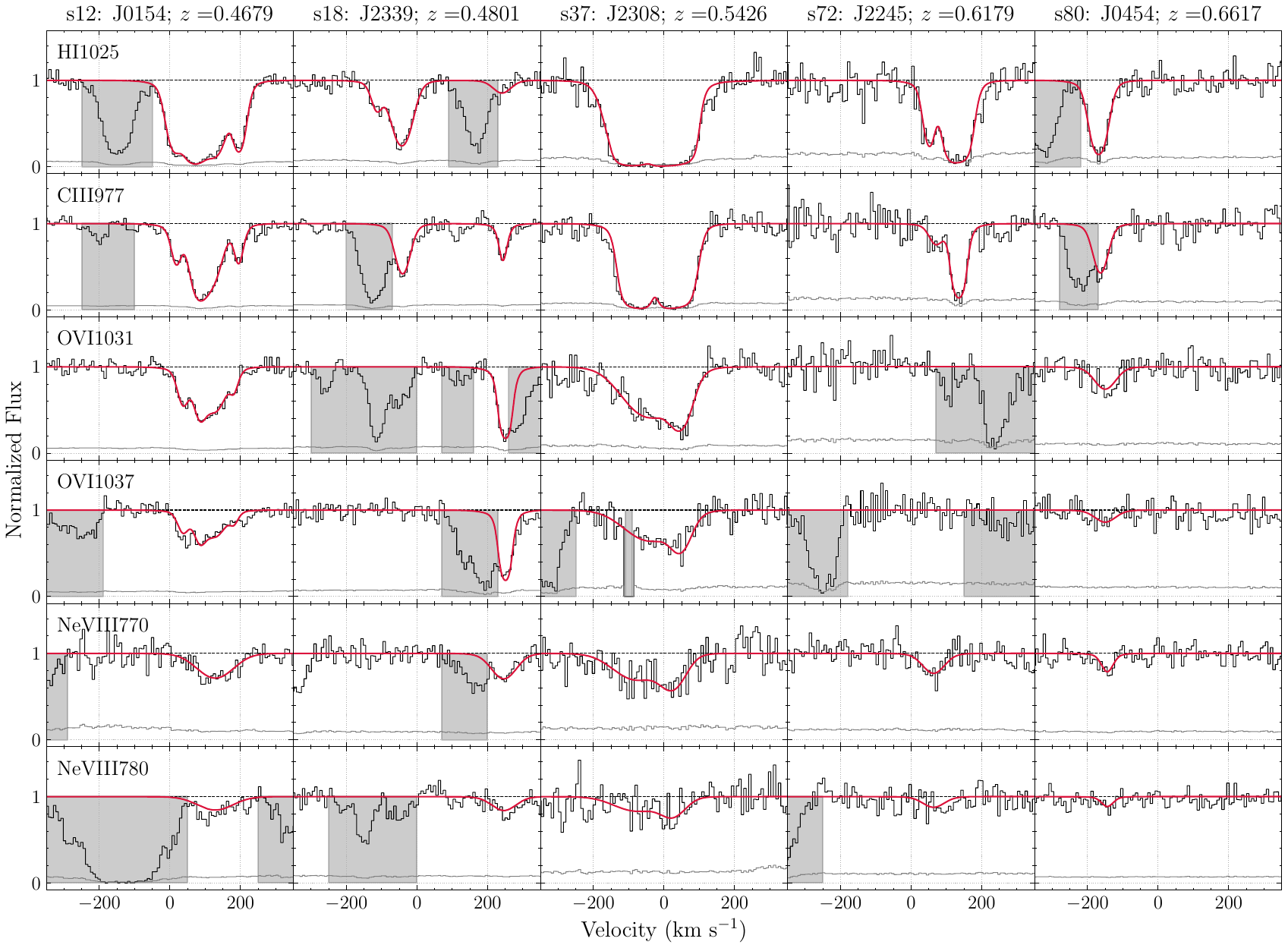}
\end{center}
\vskip -0.2in
\caption{Absorption properties of all five newly discovered \ion{Ne}{8} systems in the CUBS-midz sample, together with associated low-ionization transitions including Ly$\beta$, \ion{C}{3} $\lambda 977$, \ion{O}{6} $\lambda\lambda 1031, 1037$, and \ion{Ne}{8} $\lambda\lambda 770, 780$.
Zero velocity represents the galaxy at the smallest $\dproj/\rvir$.
The red lines are the best-fit models obtained from the Voigt profile fitting for each transition. 
The gray-shaded regions mark contaminating features or bad pixels.
}
\label{fig:ne8}
\end{figure*}

Although the nine detected \ion{Ne}{8} absorption systems exhibit a declining trend with increasing distance, the radial profile of $N_{\rm NeVIII}$ cannot be quantitatively modeled using the two-component model because of the small number of detections. 
Here, we calculate the covering fractions for the high column density \ion{Ne}{8} systems of $\log N/\cmjj \geq 14.0$ for different galaxies.
There is no robust detection of \ion{Ne}{8} in halos around dwarf galaxies, leading to $2\sigma$ upper limits in covering fractions of $36\%$ and $15\%$ within and beyond $\rvir$, respectively.
For sub-$L^*$ galaxies, the covering fraction exhibits a significant decline from $50\pm 17 \%$ within $\rvir$ to $6\pm3\%$ at $1-3\rvir$. 
In massive halos, the covering fraction is $23_{-12}^{+17}\%$ within $\rvir$.
However, the only high column density \ion{Ne}{8} is associated with a post-starburst AGN-host galaxy.
Ignoring this outlier, the covering fraction is $<41\%$ within $\rvir$.
Beyond $\rvir$, only two systems are detected with $\log N/\cmjj \approx 14.0$, which leads to a low covering fraction of $4_{-2}^{+4} \%$.
The difference in \ion{Ne}{8} covering fractions between sub-$L^*$ and massive galaxy halos suggests that sub-$L^*$ halos contain more concentrated \ion{Ne}{8}-absorbing gas, similar to \ion{O}{6}.

\section{Discussion}
\label{sec:dis}

Combining the CUBS-midz sample with previously published literature samples, we have demonstrated that galaxies with different star-formation histories, determined according to $\mstar$ and SFR, exhibit different radial profiles of \ion{O}{6} and \ion{Ne}{8} absorption properties ($N$ and $\sigma_\varv$).
In this section, we first examine the galaxy environments of the five new \ion{Ne}{8} absorbers discovered in our survey and then discuss the implications of our findings for the warm-hot CGM, including its spatial distribution,  total mass, and possible origins in halos of different masses.

\subsection{Notes on individual \ion{Ne}{8} absorbers}
\label{sec:ne8}
As described in Section \ref{sec:abs}, the detected \ion{Ne}{8} absorbers are identified based on a matched doublet ratio without significant contamination and with both doublet components detected at $\gtrsim 2 \sigma$.
All five detected \ion{Ne}{8} absorbers also exhibit low ionization species over a similar velocity range (e.g., \ion{H}{1} and \ion{C}{3}), as shown in Figure \ref{fig:ne8}.
Four of five systems have detectable \ion{O}{6} features, while one system shows a $2\sigma$ upper limit of $\log N/\cmjj< 13.9$.

\begin{figure*}
\begin{center}
\includegraphics[width=0.98\textwidth]{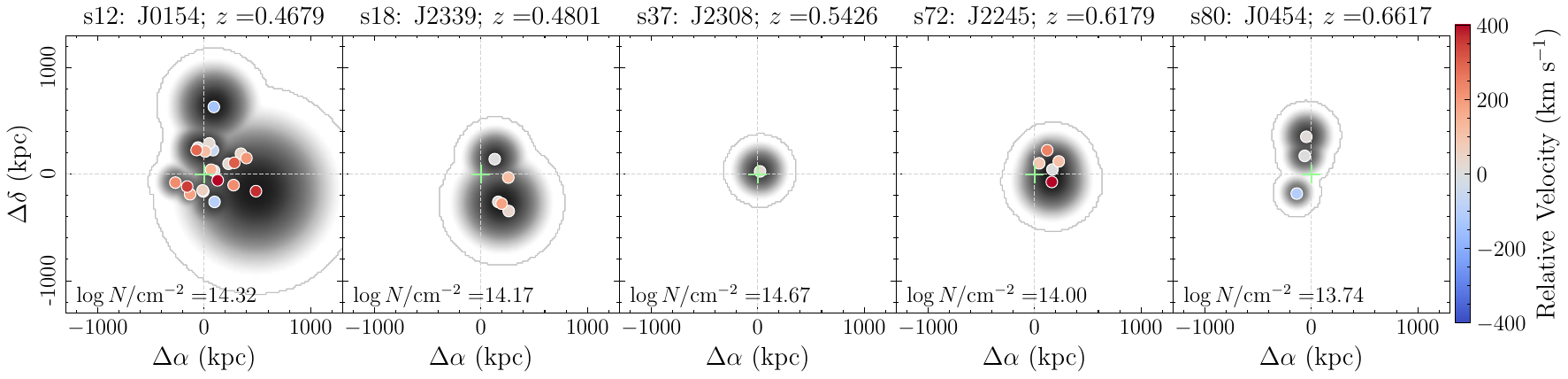}
\end{center}
\vskip -0.2in
\caption{Summary of the galactic environments of newly detected \ion{Ne}{8} systems in the CUBS-midz sample.
The \ion{Ne}{8} column densities are labeled in the lower-left corner of each panel, and the redshift at the top-right.
The QSO sightline is located at the center of each panel, marked by a green plus. 
The colors represent the velocity offsets, as shown in the color bar on the right, of individual galaxies relative to the redshift of the galaxy with the smallest $\dproj/\rvir$.
To visualize the galaxy environment, we model each associated galaxy using a 2D Gaussian, with the FWHM representing $\rvir$ and amplitude representing the inferred halo mass.
With the exception of one \ion{Ne}{8} associated with a single star-forming disk galaxy at $z=0.5426$ (middle panel), the remaining four absorbers are all found in an overdense galaxy environment. }
\label{fig:ne8_gals}
\end{figure*}

These newly discovered \ion{Ne}{8} absorbers are found in a range of galactic environments (Figure \ref{fig:ne8_gals}).
In particular, the strongest \ion{Ne}{8} of $\log N/\cmjj = 14.67\pm0.05$ is detected in the inner halo ($\dproj\approx 30$ kpc) of an isolated star-forming disk galaxy with $\log \mstar/\msun = 10.0\pm0.1$ (i.e., system s37) that produces an LLS system \citep{Chen:2020aa, Zahedy:2021aa}.
The remaining \ion{Ne}{8} absorbers are detected in galaxy groups with the number of members ranging between 5 and 21 and the total stellar mass summed over all group members ranging from $\log \mstar/\msun \approx 10.2$ to 11.4.
All of these galaxy groups have star-forming galaxies projected close to the QSO sight lines.

A comparison of the gas kinematics between \ion{Ne}{8} and \ion{O}{6} shows that \ion{Ne}{8} typically exhibits a line width that is either comparable or broader than \ion{O}{6}, indicating that at least some \ion{Ne}{8} absorbers do not share the same origin as \ion{O}{6}. 
This may be due to blending in \ion{Ne}{8} components if these components exhibit broader $b$ values.
It is worth noting that \ion{O}{6} in system s18 at $z=0.4801$ is exceptionally strong, with a $2\sigma$ lower limit of $\log N/\cmjj > 14.9$.  The doublet features cannot be explained by either an interloper \lya\ or \lyb\ line, because of a lack of corresponding higher-order Lyman series lines at shorter wavelengths. 
The aligned line centroid between \ion{O}{6} and \ion{Ne}{8} but substantially narrower \ion{O}{6} lines suggest that local ionizing radiation or non-equilibrium processes may be crucial for explaining the exceptional strength of \ion{O}{6}.

Finally, the lack of \ion{O}{6} makes system s72 a special case, which is rare in all previous \ion{Ne}{8} systems \citep[e.g.,][]{Savage:2005aa, Narayanan:2011aa, Meiring:2013aa, Burchett:2019aa}.
In this system, we are able to place a $2\sigma$ upper limit of $\log N/\cmjj < 13.9$ for the possible presence of \ion{O}{6}, while
the \ion{Ne}{8} lines are detected at a significance of $\approx 4 \sigma$ and $\approx 2\sigma$ for the strong and weak members, leading to a total significance of $\approx 5 \sigma$. 
Although individual components of low ions (e.g., \ion{H}{1} and \ion{C}{3}) are detected with a similar line centroid to the potential \ion{Ne}{8} component, the dramatically different ionization potentials make it unlikely for these low ions to share the same ionization phase with \ion{Ne}{8}.
We include this system as a \ion{Ne}{8} detection in all analyses, but it is still possible that the matched \ion{Ne}{8} doublet may be due to contaminating absorption features at different redshifts.

\subsection{Covering fraction of \ion{Ne}{8}}
\label{sec:cf}

As a tracer of the warm-hot CGM, \ion{Ne}{8} absorption has been detected in FUV spectra of background QSOs \citep[e.g.,][]{Savage:2005aa, Mulchaey:2009aa, Narayanan:2011aa, Narayanan:2012aa, Meiring:2013aa, Qu:2016aa}, although the first \ion{Ne}{8} search in a blind galaxy sample established without prior knowledge of existing features was carried out by the CASBaH survey \citep{Burchett:2019aa}.  A high covering fraction of $\kappa\approx 44\pm20\%$ was found at $d<200$ kpc for \ion{Ne}{8} of $\log N/\cmjj > 14.0$.
Taken at the face value, we would expect to detect $\sim 10$ new strong \ion{Ne}{8} absorbers with $\log N/\cmjj > 14.0$ for 23 galaxies with $\log \mstar/\msun >9$ and meaningful constraints on \ion{Ne}{8} projected within 200 kpc from QSO sightlines in the CUBS-midz sample, but only five were found.  Here we examine possible factors that may have contributed to this discrepancy.

The CASBaH galaxies span a range in \mstar\ from $\log \mstar/\msun \approx9.0$ to 11.3 with a median of $10.1$ and a standard deviation of $0.6$ dex.  The lowest-mass galaxies presented by the CASBaH are still more than an order of magnitude more massive than the low-mass galaxies in the CUBS-midz sample (see e.g., Figure \ref{fig:sample}).  Therefore, the first step toward a systematic comparison is to establish a comparable galaxy sample.
To this end, we select all galaxies with $\log \mstar/\msun > 9.0$ in the CUBS-midz sample, which has a median of $10.3$ and a standard deviation of 0.7 dex.

\begin{figure*}[t]
\begin{center}
\includegraphics[width=0.95\textwidth]{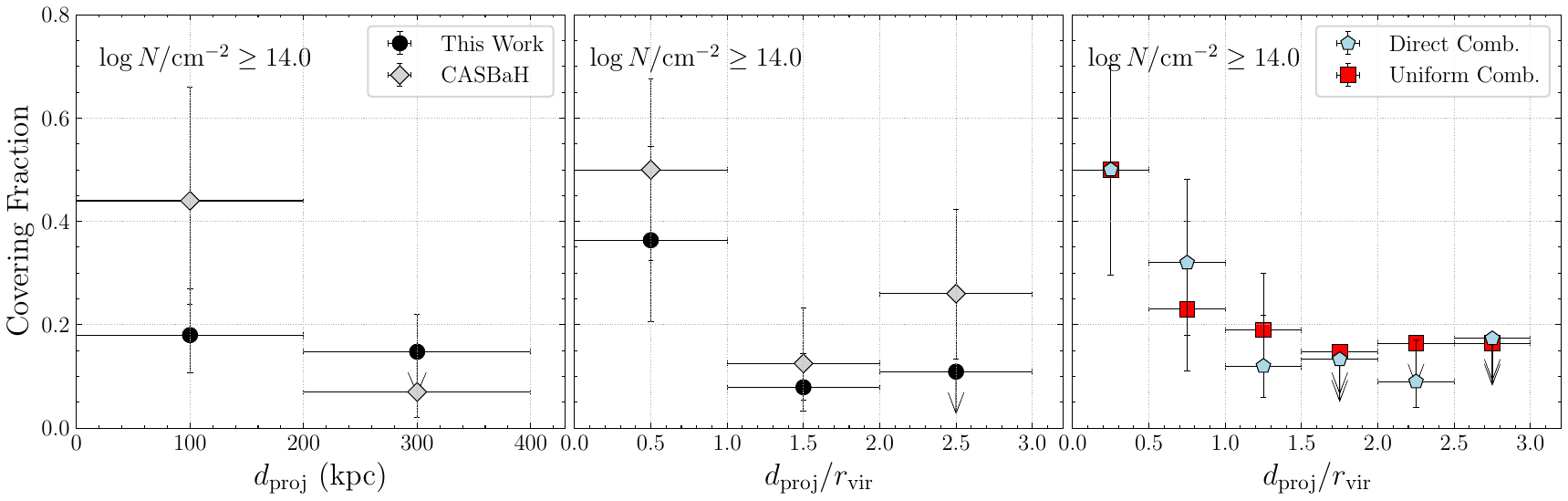}
\end{center}
\vskip -0.2in
\caption{Covering fractions of \ion{Ne}{8} absorption features with $\log N/\cmjj\geq 14.0$ as functions $\dproj$ (the left panel) and $\dvir$ (middle and right panels).
Here, we filter the stellar mass $\log \mstar/\msun>9.0$ of galaxies in the CUBS program to match with the CASBaH sample (see text for details).
In the left two panels, the CUBS sample (black circles) and the CASBaH sample (gray diamonds) are presented separately.
Although the results from the two samples are consistent within the 1-$\sigma$ uncertainties, the covering fractions estimated using the CUBS sample are consistently lower than the CASBaH measurements.
In the right panel, we show the covering fractions at $\log N/\cmjj \geq 14.0$, combining both samples.
We report two sets of combined covering fraction measurements.  The first is calculated including all \ion{Ne}{8} reported by the CASBaH team (light blue pentagons), while the second includes only absorbers that satisfy the same set of criteria as those from CUBS (red squares; see Section \ref{sec:ms}).
This uniform, combined sample results in conservative measurements of the \ion{Ne}{8} gas covering fraction, due to uncertainties in \ion{Ne}{8} identifications. 
}
\label{fig:cf}
\end{figure*}

Given the challenge of robustly identifying a \ion{Ne}{8} absorber in the presence of numerous interlopers, we also adopt a conservative column density threshold of $\log N_0/\cmjj = 14.0$ for computing the covering fraction.
To obtain accurate constraints for the gas covering fraction, particularly in the event of a small sample, it is critical to first establish a uniform sample of absorption spectra that offers a consistent sensitivity limit for detecting both strong and weak absorbers.  While strong absorbers can be detected in both low- and high-$S/N$ data, weak absorbers can only be detected in high-$S/N$ spectra.
Including low-$S/N$ sightlines based on the presence of a strong absorber, instead of matching sensitivities, would lead to an overestimate of the gas covering fraction.
For this reason, we estimate a limiting column density in continuum regions around all detections to identify and exclude those low-$S/N$ sight lines even when a strong absorber is reported.
This exercise leads to 57 and 26 galaxies/galaxy groups with sufficient sensitivities in the background QSO spectrum to detect \ion{Ne}{8} absorbers of $\log N_{\rm NeVIII}/\cmjj > 14$ in the CUBS and CASBaH surveys, respectively, projected within 3$\rvir$.

In Figure \ref{fig:cf}, covering fractions are shown as functions of both $\dproj$ and $\dvir$.
We adopt all reported \ion{Ne}{8} and their $\log N_{\rm NeVIII}/\cmjj$ from \citet{Burchett:2019aa} for the panel on $\rvir$, but compute the covering fraction versus $\dvir$ ourselves, because these are not reported in \citet{Burchett:2019aa}.
All covering fractions obtained from the CUBS and CASBaH samples are consistent with each other within $1 \sigma$ at all radii, showing significant declines beyond 200 kpc or $\rvir$.
However, the CUBS sample prefers lower covering fractions than the CASBaH survey.
Specifically, the CUBS sample exhibits a covering fraction of $18_{-7}^{+9}\%$ for \ion{Ne}{8} absorbers with $\log N/\cmjj > 14.0$, while the CASBaH sample prefers $44_{-20}^{+22}\%$ within 200 kpc.
Within $\rvir$, the CUBS and CASBaH samples show $36_{-16}^{+18}\%$ and $50\pm17\%$, respectively.
Here, the covering fractions are higher than the values in Figure \ref{fig:N_radial}, because non-detections associated with low-mass galaxies are omitted from this comparison.

For an optimal statistical analysis, we combine the CUBS and CASBaH samples to assess how the gas covering fraction depends on $\dproj/\rvir$.  This is presented in the right panel of Figure \ref{fig:cf}.
We report two sets of measurements resulted from applying different treatments for the CASBaH sample. 
First, we calculate the combined covering fraction using all nine reported \ion{Ne}{8} absorbers from \citet{Burchett:2019aa}. The results are shown as pentagons in the right panel of Figure \ref{fig:cf}.
The measured covering fractions are summarized in Table \ref{tab:cf}
Second, we apply a uniform set of selection criteria of \ion{Ne}{8} absorbers described in Section \ref{sec:ms} for both samples and compute the gas covering fractions based on this uniform sample.
This second set of measurements represents a conservative estimate of the gas covering fraction due to the difficulties in robustly identifying \ion{Ne}{8} in the presence of interlopers.

\begin{table}[]
    \centering
    \caption{Summary of the observed \ion{Ne}{8} covering fraction (\%)}
    \begin{tabular}{lccccccc}
    \hline\hline
    & \multicolumn{6}{c}{$\dvir$} \\
     \cline{2-7}
    Sample & 0.5 & 1.0 & 1.5 & 2.0 & 2.5 & 3.0 \\
    \hline
    Direct & $50\pm20$ & $32_{-14}^{+17}$ & $12_{-6}^{+9}$ & $<13$ & $9_{-5}^{+8}$ & $<17$ \\
    Uniform & $50\pm 20$ & $23_{-12}^{+17}$ & $19_{-8}^{+11}$ & $<14$ & $<17$ & $<17$ \\
    \hline
    \end{tabular}
    \label{tab:cf}
\end{table}

In summary, our analysis shows that the covering fraction \ion{Ne}{8} declines with increasing distance out to $\approx 1.5\,\rvir$, and it may not be as high as previously thought around galaxies of $\log \mstar/\msun\approx 9.0-11.5$.

\subsection{Total mass of the warm-hot CGM probed by \ion{O}{6} and \ion{Ne}{8}}
\label{sec:mass}

\begin{figure*}
\begin{center}
\includegraphics[width=0.98\textwidth]{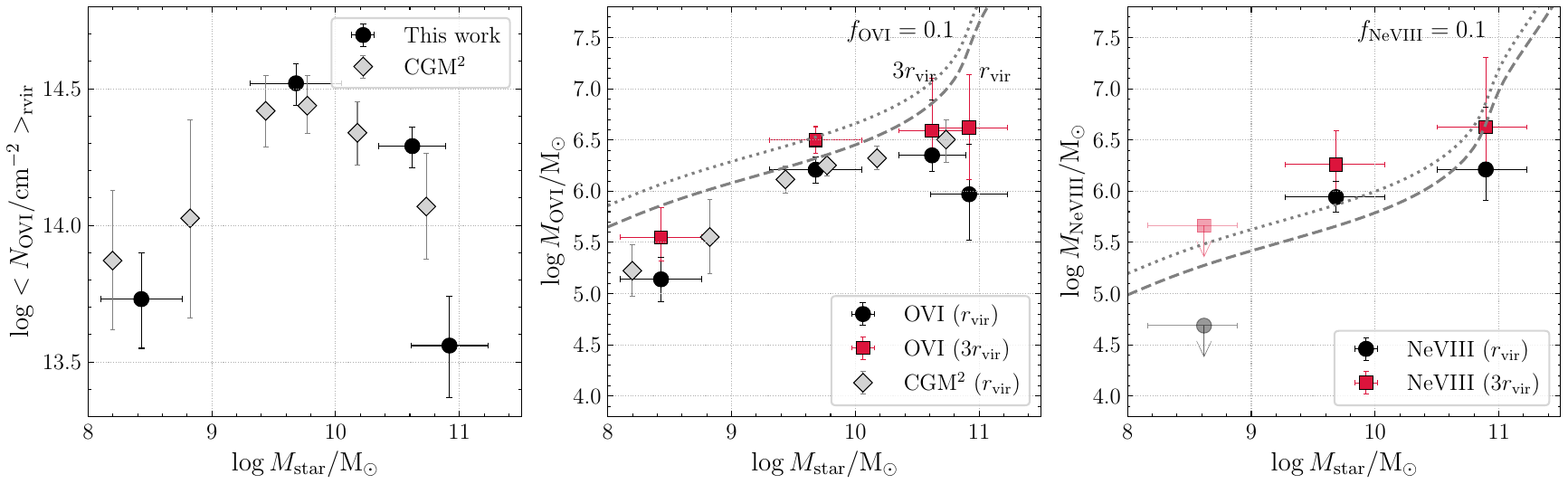}
\end{center}
\vskip -0.2in
\caption{Mass content of the warm-hot CGM probed by \ion{O}{6} and \ion{Ne}{8} versus \mstar.  Left panel: Dependence of mean $N_{\rm OVI}$ averaged within $\rvir$ on the stellar mass (black circles).
The mean $N_{\rm OVI}$ peaks at $\log \mstar/\msun \approx10$.
For comparison, we include the mean column density of star-forming galaxies adopted from \citet[][gray diamonds]{Tchernyshyov:2022aa}.
Middle panel: Total \ion{O}{6} gas mass within $\rvir$ (black circles) or $3\,\rvir$ (red squares). 
Extending to $\approx 3\,\rvir$, the total \ion{O}{6} mass is a factor of $\approx 2-4$ larger than the mass within $\rvir$ for all galaxies.
As a comparison, we include the expected total \ion{O}{6} mass within $\rvir$ (the dashed line) and $3\,\rvir$ (the dotted line) based on a model assuming a cosmic baryonic fraction, $f_b=0.156$ and an anticipated fraction of baryonic mass in \ion{O}{6}, $f_{\rm OVI}\,f_{\rm CGM}\,Z_{\rm O}/Z_{\rm O, \odot} = 0.01$. 
Adopting typical $f_{\rm CGM}$ and $Z_{\rm O}/Z_{\rm O, \odot}$, we obtain $f_{\rm OVI}\approx 0.1$ to reproduce the observed \ion{O}{6} content in the CGM of sub-$L^*$ galaxies, suggesting that \ion{O}{6}-bearing gas is the dominant phase in these galaxies (see the text for details).
Right panel: \ion{Ne}{8} mass, similar to the middle panel.
The \ion{Ne}{8} mass is calculated empirically, which may be driven by the highest column density absorbers at small $\dvir$ (see the text for details).}
\label{fig:logN_MOVI}
\end{figure*}

To evaluate the mass budget of the warm-hot CGM, we first calculate the average column density within $\rvir$ using the best-fit column density profiles and covering fractions for galaxies in different sub-samples presented in Section \ref{sec:radial2}.
Figure \ref{fig:logN_MOVI} shows the dependence of the area-weighted mean \ion{O}{6} column density with $\rvir$ on the stellar mass.
The mean column density exhibits an increase of 0.8 dex from dwarfs to sub-$L^*$ galaxies, followed by a decrease of $0.2$ dex to massive star-forming galaxies.
At $\log \mstar/\msun > 10.3$, massive quiescent galaxies on average exhibit a lower \ion{O}{6} column density in the CGM than star-forming galaxies \citep[see also][]{Zahedy:2019aa}.

We calculate the total enclosed \ion{O}{6} mass at $\dproj<\rvir$ and within the maximum radius, $\dproj=3\,\rvir$, surveyed in this study (Figure \ref{fig:logN_MOVI}).
The total \ion{O}{6} mass increases from $\dproj<\rvir$ to $\dproj<3\,\rvir$ by $\approx0.2-0.3$ dex for star-forming galaxies, while the increment is $\approx 0.6$ dex for the massive quiescent galaxies. 
This is expected because of a flatter column density profile found in these massive quiescent halos.
For the massive galaxy samples (i.e., star-forming and quiescent), the large error bars shown in Figure \ref{fig:logN_MOVI} also have significant contributions from the difference of halo sizes for different galaxies.

In Figure \ref{fig:logN_MOVI}, we also include model predictions for the \ion{O}{6} mass, following a simple halo model. For each dark matter halo, we adopt a cosmic baryonic fraction of $f_{\rm b}=0.156$ \citep{Planck:2018parameters}, a fraction of the baryonic mass in the CGM, $f_{\rm CGM}$, and a fraction of oxygen in the \ion{O}{6} ionization state, $f_{\rm OVI}$.  The expected total mass in \ion{O}{6} is then $M_{\rm OVI} = M_{\rm halo} f_{\rm b}f_{\rm CGM}/\mu \times A_{\rm O} {(\rm O/\rm H)} f_{\rm OVI}$, where $\mu=1.4$ is the atom mass per hydrogen, $A_{\rm O}=16$ is the atomic number of oxygen, and O/H is the number ratio between oxygen and hydrogen. 

In the following discussion, we use $f_{\rm CGM} f_{\rm OVI} Z_{\rm O}/Z_{\rm O, \odot}$ as an integrated normalization of the model because these quantities are degenerate with one another.  We adopt $Z_{\rm O}/Z_{\rm O, \odot}$ to represent the oxygen abundance relative to the solar value.
We find that $f_{\rm CGM} f_{\rm OVI} Z_{\rm O}/Z_{\rm O, \odot} \approx 0.01$ is required to reproduce the observed \ion{O}{6} ion mass in sub-$L^*$ galaxies as shown in Figure \ref{fig:logN_MOVI}, which display the highest mean $N_{\rm OVI}$.
A typical metallicity in the CGM is $\approx 0.3\,Z_\odot$, and the CGM mass fraction is predicted to be $0.1-0.5$ in recent simulations \citep[e.g.,][]{Wijers:2020aa}.
Adopting a typical gas-phase metallicity and $f_{\rm CGM}$, we obtain an average ionization fraction $\approx 0.05-0.3$ for \ion{O}{6}, which is comparable with the peak ionization fraction of $0.2-0.4$ in CIE or PIE models \citep[e.g.,][hereafter, \citetalias{Stern:2018aa}]{Oppenheimer:2013aa, Stern:2018aa}.
Therefore, we conclude that \ion{O}{6}-bearing gas is an important phase in these sub-$L^*$ halos. 
We also plot the model predicted mass out to $3\,\rvir$, extrapolating a Navarro-Frank-White \citep[NFW;][]{Navarro:1996aa} halo profile with a concentration of $4$ for star-forming galaxies. 
For sub-$L^*$ galaxies, the predicted \ion{O}{6} is consistent with observations, indicating \ion{O}{6} is also abundant in the outskirts of these galaxy halos.
On the other hand, the observed \ion{O}{6} mass in the CGM of dwarf galaxies or massive quiescent galaxies is significantly lower than the model prediction, suggesting that the typical ionization state in these halos is offset from \ion{O}{6} assuming $f_{\rm CGM}\approx 0.3$ \citep[also see][]{Zahedy:2019aa, Zheng:2024aa}.
However, the CGM mass fraction in dwarf galaxies may be as low as $\approx 0.1$ \citep[e.g.,][]{, Schaller:2015aa, Hafen:2019aa}, which also affects the \ion{O}{6} mass.

For \ion{Ne}{8}, no best-fit model is available because of the small sample of detections.
Therefore, the \ion{Ne}{8} mass is calculated by summing over all available empirical constraints, including both measurements and upper limits.  Specifically, we divide the radial profile into discrete bins and calculate a median \ion{Ne}{8} column density in each bin.
For non-detections, we estimate a median column density, $\langle\,N\,\rangle$ that satisfies ${\rm erf}(\langle\,N\,\rangle/N^{68})=\frac{1}{2}\,{\rm erf}(N^{95}/N^{68})$, where ${\rm erf}()$ is the error function and $N^{68}$ and $N^{95}$ represent the 68\% and 95\% upper limits, respectively. 
The empirically determined \ion{Ne}{8} masses are plotted in the right panel of Figure \ref{fig:logN_MOVI}.

Similar to \ion{O}{6}, we also plot the simple model prediction of the \ion{Ne}{8} mass assuming $f_{\rm CGM} f_{\rm NeVIII} Z_{\rm Ne}/Z_{\rm Ne,\odot} \approx 0.01$.
As a lithium-like ion, \ion{Ne}{8} shares a similar ionization fraction as \ion{O}{6}, with a peak of $f_{\rm NeVIII}\approx 0.2$ and $0.4$ in CIE and PIE, respectively \citep[e.g.][]{Oppenheimer:2013aa}.
The observed \ion{Ne}{8} mass is higher than the model prediction for the sub-$L^*$ sample.
However, we note for both the sub-$L^*$ and massive galaxy samples, the mass estimates are driven by a few high column density systems at the smallest $\dvir$, which contribute $30\%-40\%$ of the total \ion{Ne}{8} mass. The total mass estimate is, therefore, subject to large stochastic uncertainties.

We note that a systematic uncertainty in the mass budget calculation may be induced by the halo model adopted for estimating $\rvir$.
Compared to the CGM$^2$ survey, 
For sub-$L^*$ and massive star-forming galaxies, our mean \ion{O}{6} column densities are higher than the values reported in CGM$^2$ by $0.1-0.2$ dex.
This difference is mainly due to a different adopted stellar mass-halo mass relation for computing $\rvir$. 
As stated in Section \ref{sec:data}, we adopted the  \citet{Kravtsov:2018aa} relation that includes missing light correction in the total stellar mass of massive galaxies.  This leads to a smaller inferred halo mass by $\approx 0.1-0.2$ dex for a given \mstar\ at $\log \mstar/\msun\approx 9.5-10.5$, compared to the \citet{Behroozi:2019aa} model.
Using the \citet{Behroozi:2019aa} model, we would infer a higher halo mass and, as a result, a larger $\rvir$.   A larger $\rvir$ would lower the mean column density within $\rvir$ by including more non-detections at larger projected distances.

\begin{figure*}
\begin{center}
\includegraphics[width=0.98\textwidth]{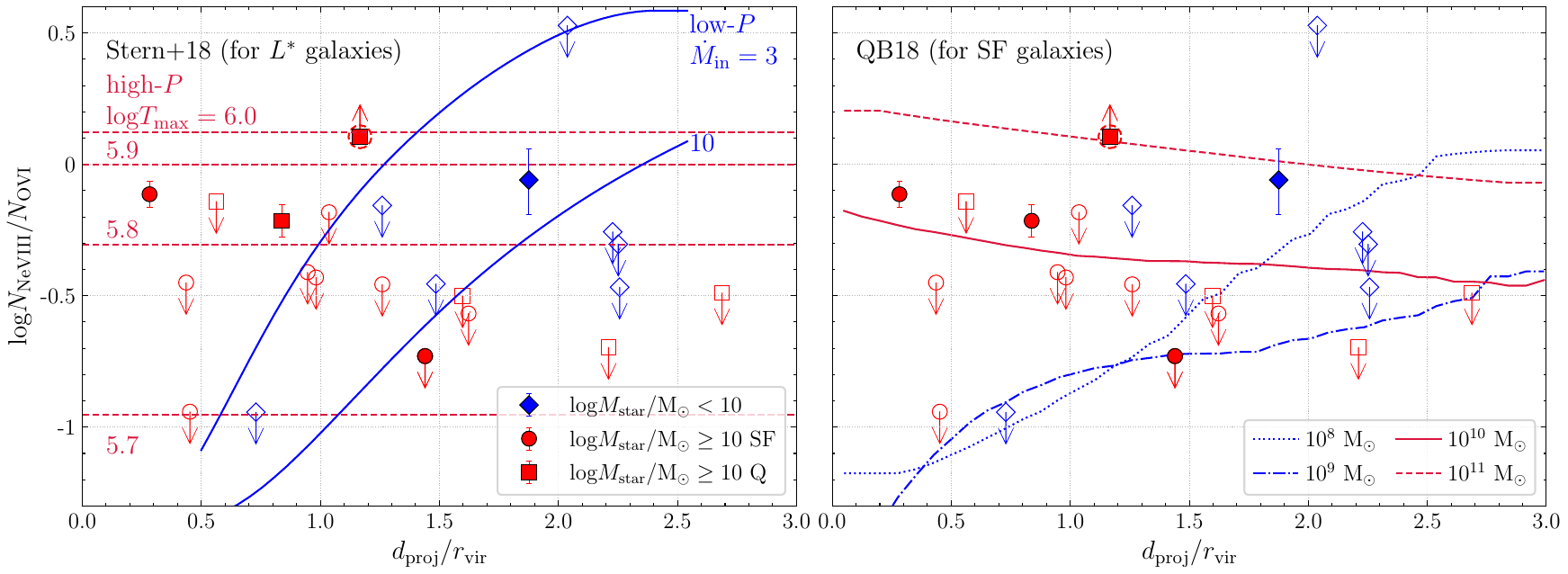}
\end{center}
\vskip -0.2in
\caption{Comparison of dependence of \neo\ on $\dvir$ between data and models.
Photoionization-dominated models predict increasing of \neo\ at larger $\dvir$ in  (i.e., low-pressure models in the left panel, \citetalias{Stern:2018aa}; and models for low-mass $\log \mstar/\msun \lesssim 9$ in the right panel, \citetalias{Qu:2018aa}).
The warm-hot gas in cooling gas or virialized gas suggests roughly constant \neo\, over different radii (i.e., high-pressure models in the left panel and models for massive $\log \mstar/\msun \gtrsim 10$ in the right panel).
For observations, only absorption systems with at least \ion{O}{6} or \ion{Ne}{8} features are included to ensure meaningful constraints. 
The small sample size limits the ability to distinguish between models.}
Open symbols represent upper limits in \ion{Ne}{8}, while \ion{O}{6} is a detection.  The filled red circle with a downward arrow is s18 at $z=0.4801$, for which both \ion{Ne}{8} and \ion{O}{6} are detected but the \ion{O}{6} is exceptionally strong and saturated.
The full sample is divided into a low-mass star-forming sample with $\log \mstar/\msun\leq 10$ (blue diamonds), massive star-forming (red circles), and quiescent samples (red squares) with $\log \mstar/\msun> 10$.
\label{fig:ratio}
\end{figure*}

\subsection{Spatial variation of the \neo\ ratio}
\label{sec:ratio}

The origin of the warm-hot CGM may be explained by different scenarios, including the virialized ambient gas, the outflowing ejection from galaxy disk, and cooling gas from the hotter phase (e.g., \citealt{Oppenheimer:2016aa, McQuinn:2018aa}, \citetalias{Stern:2018aa}, \citetalias{Qu:2018aa}, \citealt{Wijers:2024aa}).
The observed \neo\ ratio provides valuable insights into the ionization mechanisms of the warm-hot CGM and its origin.
In particular, we consider two different model scenarios: photoionized gas vs.\ radiative cooling gas in the hot halo 
(see \citetalias{Stern:2018aa} and \citetalias{Qu:2018aa}).
These models predict distinct behaviors for \neo\ in different scenarios.
However, the current sample is still too small to establish a consistent physical understanding of the ionization mechanism of these absorbers.
Here, we compare the existing sample with model predictions to shed light on future observations.

\citetalias{Stern:2018aa} introduced two scenarios to generate \ion{O}{6} and \ion{Ne}{8} in the CGM of star-forming $L^*$ galaxies.
These authors considered the possibility that the warm-hot CGM is photoionized in the low-density and low-pressure halo. 
This low-pressure scenario shows a clear increase in \neo\ with increasing projected distance
because of a declining gas density (Figure \ref{fig:ratio}).
The predicted ion ratios from this low-pressure model presented in Figure \ref{fig:ratio} are for two different mass inflow rates, $\dot{M}_{\rm in}=3$ and 10 $\msun~\rm yr^{-1}$; see \citetalias{Stern:2018aa} for detail).

\citetalias{Stern:2018aa} also considered the possibility that
the warm-hot gas arises in the high-pressure cooling flow from the hot phase \citep[also see][]{Stern:2019aa, Stern:2023aa}.
In this high-pressure scenario, the \neo\ ratio only depends on the maximum temperature of the cooling flow, which shows no radial dependence.

\citetalias{Qu:2018aa} developed a hybrid model for galaxies in different mass ranges.
For a gaseous halo around galaxies of $\mstar\approx 10^8$ to $10^{11}~\msun$, these authors attributed the warm-hot gas to cooling gas from the hot halos and accounted for photoionization due to the ultra-violet background (UVB) by adopting the ionization fractions from \citet{Oppenheimer:2013aa}.
In low-mass galaxies of $ \mstar\approx 10^8-10^9~\msun$, the virial temperature is too low to form \ion{O}{6} and \ion{Ne}{8} through collisional ionization.
Therefore, the measured warm-hot CGM is photoionized, which shows increasing \neo\ with radius.
For massive galaxies of $\mstar\gtrsim 10^{10}~\msun$, the \citetalias{Qu:2018aa} model predicts relatively constant \neo\ with radius because the warm-hot gas is mainly collisionally ionized at the peak temperature. 

Figure \ref{fig:ratio} shows comparisons between predicted and observed \neo\ ratios at different projected distances.
The observed \neo\ ratios are presented for absorbers with at least one ion detected (i.e., \ion{O}{6} or \ion{Ne}{8}).
In total, 23 individual systems have meaningful constraints on \neo, including all five newly discovered \ion{Ne}{8} absorbers (see Section \ref{sec:ne8}).

The galaxy sample is divided into three sub-samples according to $\mstar$ and SFR.
For low-mass galaxies of $\log \mstar/\msun <10$, the only measured \neo\ is at $\approx 2\rvir$.
This is consistent with the photoionization scenario, although no trend can be derived with only one detection in this mass range.
For massive galaxies of $\log \mstar/\msun \geq 10$, we further group them into star-forming and quiescent galaxy sub-samples with a SFR threshold of $0.2\,\msun\rm yr^{-1}$.
These two sub-samples exhibit similar \neo\  ratios.
For massive star-forming galaxies, two systems with both \ion{Ne}{8} and \ion{O}{6} detected show \neo\ of $\approx 0.6-1$, suggesting a temperature of $\log T/{\rm K}\approx 5.8$ in either the collisional ionization scenarios (e.g., \citealt{Savage:2005aa, Meiring:2013aa, McQuinn:2018aa}).

We note two special cases.  For s72 with an \neo\ ratio of $\gtrsim 1.2$, the associated galaxy at the smallest $\dvir$ is the most massive with $\log \mstar/\msun \approx 10.6$ among all five detected \ion{Ne}{8} absorbers.  This is indeed expected from the cooling flow model, in which massive galaxies should produce the highest \neo\ ratio \citepalias{Qu:2018aa}.  For s18 with
an \neo\ ratio of $\lesssim 0.2$, the observed \ion{O}{6} is exceptionally strong and narrow, suggesting a photoionized origin or a non-equilibrium condition.
It is therefore likely that this system is significantly affected by local radiation due to nearby galaxies (see examples in \citealt{Zahedy:2021aa}; \citetalias{CUBSVI}).

Given the distinct predictions for \neo\ from different model scenarios, our sample at the moment is still too small to establish a consistent physical understanding of the origin of these absorbers. In particular, existing data do not provide a sufficiently large dynamic range in the \neo\ ratio to discriminate different models.
Improving the detection limit of \ion{Ne}{8} with high S/N ratio spectra is necessary to strengthen these model comparisons.

\section{Summary}
\label{sec:summary}

In this work, we establish the CUBS-midz sample of $0.4\lesssim z \lesssim 0.7$ galaxies with small projected distances to the 15 CUBS background QSOs (Figure \ref{fig:sample}), aiming to study the warm-hot CGM probed by the \ion{O}{6} and \ion{Ne}{8} doublets.
The new sample contains 50 individual galaxies and 53 galaxy groups with 2 to 21 member galaxies.
In this sample, 30 \ion{O}{6} and 5 \ion{Ne}{8} absorbers are detected.
Combining the CUBS-midz sample with literature samples, we investigate the radial dependence of line-of-sight absorption properties (e.g., the column density and velocity dispersion) on galaxy stellar masses and SFRs.
Our major results are summarized below.
\begin{itemize}
    \item Detected \ion{O}{6} and \ion{Ne}{8} absorption features exhibit column densities of $\log N/\cmjj \approx 13.5- 15.0$, while the limiting column densities are $\log N/\cmjj \approx13.7$ and $14.0$, respectively, for a large fraction of non-detections.
    We calculate the velocity dispersion of detected \ion{O}{6} and \ion{Ne}{8} features, spanning from $\approx 5 \kms$ to $ 120 \kms$ (Figure \ref{fig:abs_sample}).
    In particular, we consider features with $\sigma_\varv \gtrsim 40\kms$ as broad features, tracing the kinematics in the halo instead of the internal velocity dispersion in individual clouds. 
    \item We examine the correlations between the \ion{O}{6} and \ion{Ne}{8} properties and different nearby galaxies with the smallest $\dproj$, smallest $\dvir$, and highest $\mstar$, using the 53 sight lines probing the overdense galaxy environments with multiple nearby galaxies (Figures \ref{fig:gal_comp} and \ref{fig:vel_sample}).
    The observed \ion{O}{6} column density and kinematics are most correlated with the galaxy with the smallest $\dproj/\rvir$, which is therefore taken as the host galaxy of absorption features in group environments.
    \item The detected \ion{O}{6} and \ion{Ne}{8} absorption features exhibit bulk velocities within $\approx 200\kms$ relative to the associated galaxies (Figure \ref{fig:vel_sample}).
    Normalized by the escape velocity, the bulk velocity distribution can be modeled by a Gaussian function with a standard deviation of $\sigma \approx 0.40$. 
    \item 
    For both \ion{O}{6} and \ion{Ne}{8}, particularly high column density and broad absorption features are detected around galaxies with $\mstar/\msun \approx 10$ (Figures \ref{fig:N_ms} and \ref{fig:b_ms}).
    \item For \ion{O}{6}, we divide the CUBS-midz sample into four sub-samples based on $\mstar$ and SFR: the dwarf, sub-$L^*$, massive star-forming, and massive quiescent galaxy samples.
    Combined with literature samples, the sub-$L^*$ and massive star-forming galaxy samples show substantial radial declines of column density with a power law slope of $\approx 0.8$, while the massive quiescent galaxy sample exhibits flatter radial profiles (Figure \ref{fig:nb_rad_gals}).
    \item The covering fraction of \ion{O}{6} is high ($\approx 90-100\%$) for sub-$L^*$ and massive star-forming galaxies within the virial radius, and exhibits a sharp decline at $1-2\rvir$.
    The massive quiescent galaxies exhibit a roughly constant covering fraction of $\approx 30\%$ out to $3\rvir$ (Figure \ref{fig:nb_rad_gals}).
    The joint observations of flat covering fraction and column density profiles suggest less concentrated \ion{O}{6}-bearing gas in the CGM of quiescent galaxies.
    \item For \ion{Ne}{8}, we divide the sample into three sub-samples based only on $\mstar$ considering the small sample of \ion{Ne}{8} detections.
    Within $\rvir$, the sub-$L^*$ galaxy sample exhibits a higher covering fraction of high $N_{\rm NeVIII}$ systems with $\log N_{\rm NeVIII}/\cmjj \geq 14.0$ (Figure \ref{fig:radial_ne8}).
    \item We compare the covering fraction of \ion{Ne}{8} with the CASBaH sample for galaxies with $\log \mstar/\msun \approx 9 -11.5$, showing that the CUBS sample exhibits relatively lower covering fractions over all radii for a limiting $\log N_{\rm NeVIII}/\cmjj=14.0$ (Figure \ref{fig:cf} and Table \ref{tab:cf}).
    \item The warm-hot CGM probed by \ion{O}{6} and \ion{Ne}{8} dominates the CGM of sub-$L^*$ galaxies, exhibiting the highest area-weighted mean column density within $\rvir$ (Figure \ref{fig:logN_MOVI}).
    Adopting a typical metallicity and CGM mass fraction, we show that ionization fractions of \ion{O}{6} and \ion{Ne}{8} are comparable with the peak ionization fraction in CIE or PIE models, suggesting the warm-hot CGM probed by \ion{O}{6} and \ion{Ne}{8} is the dominant phase in halos of sub-$L^*$ galaxies.
    
\end{itemize}

\section*{Acknowledgements}
We thank the anonymous referee for a careful review and valuable suggestions that significantly improved our work.  
We thank Jonathan Stern for sharing their models of \neo\, ratio in $L^*$ galaxies.
ZQ acknowledges partial support from HST-GO-15163.001A, NSF AST-1715692 grants, and NASA ADAP grant 80NSSC22K0481.
HWC and MCC acknowledge partial support from HST-GO-15163.001A and NSF AST-1715692 grants.
SDJ acknowledges partial support from HST-GO-15280.009A.
GCR acknowledges partial support from HST-GO-15163.015A.
FSZ acknowledges the support of a Carnegie Fellowship from the Observatories of the Carnegie Institution for Science.
DD acknowledges the support of the John A. Lyons Fellowship from MIT’s Office of Graduate Education.
EB acknowledges partial support from NASA under award No. 80GSFC21M0002.
SC gratefully acknowledges support from the European Research Council (ERC) under the European Union’s Horizon 2020 research and innovation programme grant agreement No 864361.
CAFG was supported by NSF through grants AST-2108230, AST-2307327, and CAREER award AST-1652522; by NASA through grants 17-ATP17-0067 and 21-ATP21-0036; by STScI through grants HST-GO-16730.016-A and
JWSTAR-03252.001-A; and by CXO through grant TM2-23005X.
JIL is supported by the Eric and Wendy Schmidt AI in Science Postdoctoral Fellowship, a Schmidt Futures program.
This work is based on observations made with ESO Telescopes at the Paranal Observatory under program ID 0104.A-0147(A), observations made with the 6.5m Magellan Telescopes located at Las Campanas Observatory, and spectroscopic data gathered under the HST-GO-15163.01A program using the NASA/ESA Hubble Space Telescope operated by the Space Telescope Science Institute and the Association of Universities for Research in Astronomy, Inc., under NASA contract NAS 5-26555.
This research has made use of the services of the ESO Science Archive Facility and the Astrophysics Data Service (ADS)\footnote{\url{https://ui.adsabs.harvard.edu}}. The analysis in this work was greatly facilitated by the following \texttt{python} packages:  \texttt{Numpy} \citep{Numpy2020}, \texttt{Scipy} \citep{Scipy}, \texttt{Astropy} \citep{astropy:2013,astropy:2018}, \texttt{Matplotlib} \citep{Matplotlib}, and \texttt{emcee} \citep{Foreman-Mackey:2013aa}. 






\bibliography{ms}{}
\bibliographystyle{aasjournal}

\clearpage

\newpage

\appendix
\section{Connecting \ion{Ne}{8} absorbers to galaxies in the presence of neighbors}
In Figure \ref{fig:gal_comp_ne8}, we show the $N_{\rm NeVIII}$ dependence on $\dproj$ and $\dvir$ adopting the three galaxy associations described in Section \ref{sec:host}.
Unlike \ion{O}{6}, the small number of \ion{Ne}{8} detections leads to no clear trend.

\renewcommand\thefigure{A.\arabic{figure}}
\renewcommand\thetable{B.\arabic{table}}    

\setcounter{figure}{0}  
\setcounter{table}{0}

\begin{figure}
\begin{center}
\includegraphics[width=0.45\textwidth]{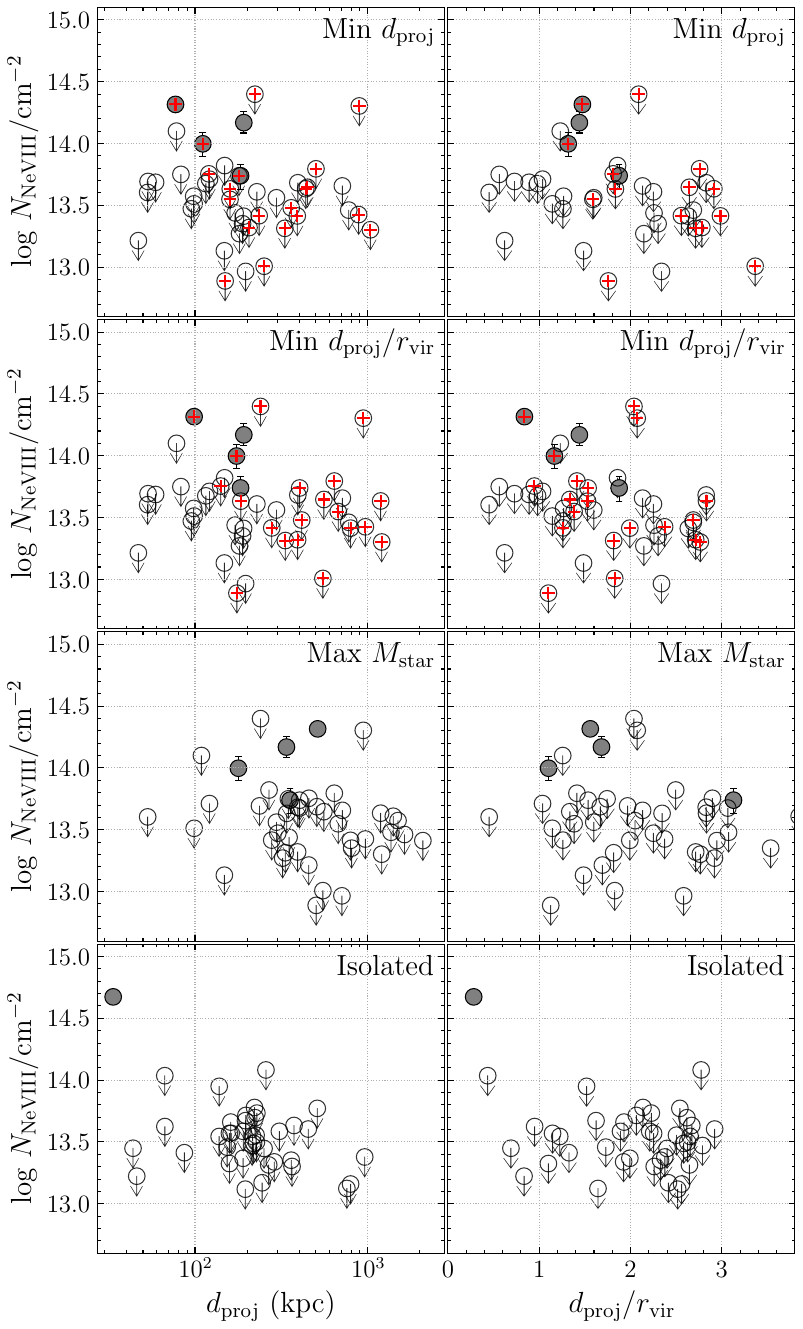}
\end{center}
\vskip -0.2in
\caption{Similar to Figure \ref{fig:gal_comp} but for the observed constraints on the \ion{Ne}{8} column densities. 
Unlike \ion{O}{6}, few galaxies/galaxy groups display detectable \ion{Ne}{8} absorption.  No clear trend can be established in any panels.  }
\label{fig:gal_comp_ne8}
\end{figure}

\section{Summary of galaxy and absorber properties}
The galaxy and absorber properties used in this study are summarized in Table \ref{tab:sample}.

\startlongtable
\begin{longrotatetable}
\begin{deluxetable*}{lcccccccccccccc}
\tablecaption{Summary of galaxy and absorber properties \label{tab:sample}}
\tablehead{
\colhead{QSO} & \colhead{ID}& \colhead{$N_{\rm gal}$} & \colhead{$z_{\rm gal}$} & \colhead{$\log M_{\rm star}$} &
\colhead{SFR} & \colhead{$d_{\rm proj}$} & \colhead{$d_{\rm proj}$} & \colhead{$\log\psi/G$} & \colhead{$\log N_{\rm OVI}$} & \colhead{$\sigma_{\rm OVI}$} & \colhead{$\varv_{\rm OVI}$} & \colhead{$\log N_{\rm NeVIII}$} & \colhead{$\sigma_{\rm NeVIII}$} & \colhead{$\varv_{\rm NeVIII}$} \\
\colhead{} & \colhead{}& \colhead{} & \colhead{} & \colhead{$/{\rm M_\odot}$} &
\colhead{$\rm M_\odot~yr^{-1}$} & \colhead{kpc} & \colhead{$/r_{\rm vir}$} & \colhead{$/(\rm M_\odot/kpc)$} & \colhead{$/{\rm cm^{-2}}$} & \colhead{$\kms$} & \colhead{$\kms$} & \colhead{$/{\rm cm^{-2}}$} & \colhead{$\kms$} & \colhead{$\kms$}
}

\startdata
J0454 & s1   &  4 & 0.4315 & $10.4\pm 0.1 $ & $<0.1$ & 82 & 0.56 & 10.1 & $13.89_{-0.03}^{+0.03}$ & $16.4_{-1.9}^{+2.1}$ & $-72.4_{-1.9}^{+2.1}$ & $<13.8$ & ... & ...\\
J0357 & s2   &  1 & 0.4358 & $10.5\pm 0.1 $ & $0.6\pm 0.2 $ & 66 & 0.44 & 10.0 & $14.49_{-0.02}^{+0.02}$ & $80.8_{-6.9}^{+6.3}$ & $-63.6_{-6.3}^{+5.8}$ & $<14.0$ & ... & ...\\
J0154 & s3   &  1 & 0.4393 & $10.5\pm 0.1 $ & $0.8\pm 0.4 $ & 452 & 2.92 & 9.2 & $<12.8$ & ... & ... & $<13.6$ & ... & ...\\
J0028 & s4   &  1 & 0.4420 & $9.2\pm 0.1 $ & $0.3\pm 0.2 $ & 197 & 2.06 & 8.9 & $<12.9$ & ... & ... & $<13.7$ & ... & ...\\
J0357 & s5   &  1 & 0.4490 & $8.2\pm 0.1 $ & $0.2\pm 0.1 $ & 66 & 0.95 & 9.0 & ... & ... & ... & $<13.6$ & ... & ...\\
J2135 & s6   &  3 & 0.4499 & $9.6\pm 0.1 $ & $2.1\pm 0.3 $ & 142 & 1.33 & 9.9 & $14.35_{-0.02}^{+0.02}$ & $18.0_{-1.0}^{+1.2}$ & $6.4_{-1.0}^{+1.0}$ & ... & ... & ...\\
J0119 & s7   &  5 & 0.4519 & $10.5\pm 0.1 $ & $1.3\pm 0.3 $ & 140 & 0.95 & 9.9 & $14.16_{-0.03}^{+0.02}$ & $22.1_{-1.9}^{+2.4}$ & $-25.0_{-1.6}^{+1.6}$ & $<13.8$ & ... & ...\\
J0154 & s8   &  1 & 0.4551 & $7.9\pm 0.4 $ & $0.2\pm 0.1 $ & 43 & 0.69 & 9.0 & $<13.0$ & ... & ... & $<13.4$ & ... & ...\\
J0119 & s9   &  1 & 0.4668 & $8.6\pm 0.2 $ & $0.3\pm 0.2 $ & 215 & 2.65 & 8.7 & $<13.0$ & ... & ... & $<13.5$ & ... & ...\\
J0111 & s10  &  1 & 0.4669 & $8.3\pm 0.3 $ & $0.1\pm 0.1 $ & 91 & 1.28 & 8.9 & $13.86_{-0.06}^{+0.05}$ & $30.8_{-4.3}^{+4.8}$ & $58.5_{-2.1}^{+1.1}$ & ... & ... & ...\\
J0114 & s11  &  3 & 0.4671 & $9.9_{-0.2}^{+0.1}$ & $1.4\pm 0.3 $ & 239 & 2.04 & 9.4 & $13.87_{-0.11}^{+0.08}$ & $33.0_{-7.7}^{+12.2}$ & $-13.3_{-3.2}^{+6.2}$ & $<14.4$ & ... & ...\\
J0154 & s12  & 21 & 0.4679 & $10.0\pm 0.1 $ & $0.3\pm 0.1 $ & 98 & 0.84 & 10.5 & $14.53_{-0.02}^{+0.02}$ & $44.9_{-1.3}^{+1.2}$ & $99.6_{-1.6}^{+2.0}$ & $14.32_{-0.06}^{+0.06}$ & $46.9_{-7.4}^{+7.5}$ & $127.9_{-1.1}^{+0.2}$\\
J0110 & s13  &  2 & 0.4725 & $10.0\pm 0.1 $ & $0.5\pm 0.2 $ & 53 & 0.45 & 9.8 & $14.55_{-0.02}^{+0.02}$ & $95.1_{-1.8}^{+1.6}$ & $75.2_{-3.1}^{+3.1}$ & $<13.6$ & ... & ...\\
J0119 & s14  &  1 & 0.4727 & $9.6\pm 0.1 $ & $0.9\pm 0.2 $ & 250 & 2.40 & 8.9 & $<13.2$ & ... & ... & $<13.4$ & ... & ...\\
J0028 & s15  &  4 & 0.4743 & $8.4\pm 0.2 $ & $0.2\pm 0.1 $ & 95 & 1.26 & 9.5 & $13.63_{-0.07}^{+0.07}$ & $21.8_{-4.5}^{+5.4}$ & $-15.5_{-5.0}^{+4.8}$ & $<13.5$ & ... & ...\\
J2308 & s16  &  1 & 0.4781 & $10.0\pm 0.1 $ & $1.0\pm 0.3 $ & 195 & 1.62 & 9.2 & $14.24_{-0.03}^{+0.03}$ & $19.0_{-2.2}^{+2.2}$ & $-23.5_{-2.3}^{+2.2}$ & $<13.7$ & ... & ...\\
J0357 & s17  &  6 & 0.4786 & $8.4\pm 0.1 $ & $0.1\pm 0.1 $ & 47 & 0.62 & 10.3 & $<12.9$ & ... & ... & $<13.2$ & ... & ...\\
J2339 & s18  &  5 & 0.4801 & $10.3\pm 0.1 $ & $0.4\pm 0.2 $ & 191 & 1.44 & 9.9 & $>14.9$ & $6.3_{-1.1}^{+2.1}$ & $249.6_{-1.1}^{+1.1}$ & $14.17_{-0.08}^{+0.09}$ & $32.0_{-9.1}^{+14.8}$ & $244.5_{-8.9}^{+5.5}$\\
J0154 & s19  &  1 & 0.4813 & $11.3\pm 0.1 $ & $<0.1$ & 756 & 1.64 & 10.4 & $<13.2$ & ... & ... & $<13.1$ & ... & ...\\
J2245 & s20  &  1 & 0.4894 & $9.8\pm 0.1 $ & $0.1\pm 0.1 $ & 137 & 1.22 & 9.3 & $<13.5$ & ... & ... & $<13.5$ & ... & ...\\
J0333 & s21  &  1 & 0.5010 & $10.6\pm 0.1 $ & $0.4\pm 0.2 $ & 360 & 2.33 & 9.3 & $<12.7$ & ... & ... & $<13.4$ & ... & ...\\
J0154 & s22  &  2 & 0.5072 & $8.4\pm 0.4 $ & $0.1\pm 0.1 $ & 170 & 2.26 & 8.9 & $13.91_{-0.04}^{+0.04}$ & $19.1_{-2.7}^{+3.1}$ & $39.0_{-2.5}^{+2.9}$ & $<13.4$ & ... & ...\\
J2135 & s23  &  1 & 0.5083 & $8.7_{-0.3}^{+0.2}$ & $0.2\pm 0.1 $ & 65 & 0.81 & 9.2 & $13.90_{-0.05}^{+0.05}$ & $33.2_{-6.6}^{+6.6}$ & $-2.0_{-5.0}^{+5.4}$ & ... & ... & ...\\
J0111 & s24  &  1 & 0.5140 & $8.7\pm 0.2 $ & $0.3\pm 0.2 $ & 238 & 2.94 & 8.6 & $<12.7$ & ... & ... & ... & ... & ...\\
J0028 & s25  &  3 & 0.5198 & $10.6\pm 0.1 $ & $5.3\pm 0.8 $ & 414 & 2.69 & 10.2 & $<12.8$ & ... & ... & $<13.5$ & ... & ...\\
J0248 & s26  &  1 & 0.5205 & $9.1\pm 0.1 $ & $<0.1$ & 137 & 1.52 & 9.0 & $<13.2$ & ... & ... & $<13.9$ & ... & ...\\
J0420 & s27  &  7 & 0.5247 & $11.3\pm 0.1 $ & $<0.2$ & 1203 & 2.76 & 10.2 & $<13.3$ & ... & ... & $<13.3$ & ... & ...\\
J0248 & s28  &  3 & 0.5265 & $11.3\pm 0.1 $ & $<0.1$ & 940 & 2.07 & 10.3 & $<13.2$ & ... & ... & $<14.3$ & ... & ...\\
J2339 & s29  &  1 & 0.5277 & $8.8\pm 0.1 $ & $0.4\pm 0.2 $ & 218 & 2.58 & 8.7 & $<13.6$ & ... & ... & $<13.5$ & ... & ...\\
J0357 & s30  &  1 & 0.5306 & $9.5\pm 0.1 $ & $1.1\pm 0.3 $ & 244 & 2.41 & 8.9 & $<13.3$ & ... & ... & $<13.2$ & ... & ...\\
J2245 & s31  &  9 & 0.5338 & $8.1\pm 0.4 $ & $0.1\pm 0.1 $ & 59 & 0.89 & 10.3 & $<13.0$ & ... & ... & $<13.7$ & ... & ...\\
J2308 & s32  &  1 & 0.5364 & $9.6\pm 0.1 $ & $2.2\pm 0.3 $ & 228 & 2.23 & 9.0 & $13.99_{-0.05}^{+0.06}$ & $10.5_{-2.9}^{+4.0}$ & $-31.9_{-1.9}^{+2.0}$ & $<13.7$ & ... & ...\\
J0111 & s33  &  1 & 0.5367 & $9.4\pm 0.1 $ & $0.6\pm 0.2 $ & 222 & 2.31 & 8.9 & $<13.2$ & ... & ... & ... & ... & ...\\
J0110 & s34  & 10 & 0.5406 & $11.2\pm 0.1 $ & $0.6\pm 0.3 $ & 364 & 0.88 & 10.7 & $13.80_{-0.08}^{+0.06}$ & $14.8_{-3.6}^{+3.9}$ & $138.4_{-3.2}^{+3.1}$ & ... & ... & ...\\
J0333 & s35  &  4 & 0.5410 & $8.8\pm 0.2 $ & $0.1\pm 0.1 $ & 196 & 2.34 & 9.9 & $<13.1$ & ... & ... & $<13.0$ & ... & ...\\
J0028 & s36  &  1 & 0.5421 & $10.6\pm 0.1 $ & $0.2\pm 0.1 $ & 364 & 2.26 & 9.4 & $<13.6$ & ... & ... & $<13.3$ & ... & ...\\
J2308 & s37  &  1 & 0.5426 & $10.0\pm 0.1 $ & $1.9\pm 0.4 $ & 33 & 0.28 & 10.0 & $14.79_{-0.02}^{+0.02}$ & $67.7_{-2.5}^{+2.4}$ & $-10.7_{-3.2}^{+2.8}$ & $14.67_{-0.05}^{+0.05}$ & $69.4_{-6.1}^{+5.7}$ & $-31.5_{-7.9}^{+8.7}$\\
J0248 & s38  &  1 & 0.5451 & $9.2\pm 0.2 $ & $0.5\pm 0.1 $ & 257 & 2.77 & 8.8 & $<13.3$ & ... & ... & $<14.1$ & ... & ...\\
J0110 & s39  &  1 & 0.5462 & $11.0\pm 0.1 $ & $<0.1$ & 794 & 2.56 & 9.9 & $<13.3$ & ... & ... & $<13.2$ & ... & ...\\
J2135 & s40  &  1 & 0.5465 & $9.7\pm 0.1 $ & $0.8\pm 0.2 $ & 176 & 1.67 & 9.1 & $<13.1$ & ... & ... & ... & ... & ...\\
J0154 & s41  & 18 & 0.5491 & $10.6\pm 0.1 $ & $0.3\pm 0.1 $ & 174 & 1.10 & 10.9 & $<13.2$ & ... & ... & $<12.9$ & ... & ...\\
J0357 & s42  &  5 & 0.5497 & $8.7\pm 0.2 $ & $0.3\pm 0.1 $ & 148 & 1.86 & 9.5 & $<13.3$ & ... & ... & $<13.8$ & ... & ...\\
J0420 & s43  &  5 & 0.5533 & $11.0\pm 0.1 $ & $0.3\pm 0.2 $ & 403 & 1.54 & 10.0 & $<13.4$ & ... & ... & $<13.7$ & ... & ...\\
J0119 & s44  &  1 & 0.5534 & $9.5\pm 0.1 $ & $8.6\pm 1.3 $ & 190 & 1.89 & 9.0 & $<13.5$ & ... & ... & $<13.6$ & ... & ...\\
J0357 & s45  &  2 & 0.5561 & $8.9\pm 0.1 $ & $1.4\pm 0.2 $ & 98 & 1.14 & 9.2 & $<13.5$ & ... & ... & $<13.5$ & ... & ...\\
J0420 & s46  &  7 & 0.5600 & $11.2\pm 0.1 $ & $<0.1$ & 794 & 1.99 & 10.4 & $<13.2$ & ... & ... & $<13.4$ & ... & ...\\
J0111 & s47  &  1 & 0.5630 & $9.2\pm 0.1 $ & ... & 159 & 1.72 & 9.0 & $<13.1$ & ... & ... & ... & ... & ...\\
J0454 & s48  &  1 & 0.5681 & $8.8\pm 0.2 $ & ... & 161 & 1.93 & 8.9 & $<13.5$ & ... & ... & $<13.7$ & ... & ...\\
J2135 & s49  &  2 & 0.5717 & $8.6\pm 0.1 $ & ... & 92 & 1.19 & 9.2 & $14.18_{-0.04}^{+0.04}$ & $17.4_{-2.0}^{+2.3}$ & $-51.0_{-2.2}^{+2.1}$ & ... & ... & ...\\
J2339 & s50  &  4 & 0.5728 & $8.4\pm 0.1 $ & ... & 53 & 0.73 & 9.6 & $14.64_{-0.03}^{+0.04}$ & $36.1_{-2.9}^{+4.8}$ & $-6.5_{-3.5}^{+2.9}$ & $<13.7$ & ... & ...\\
J0111 & s51  &  8 & 0.5766 & $10.6\pm 0.1 $ & $<0.1$ & 41 & 0.26 & 10.5 & $<13.5$ & ... & ... & ... & ... & ...\\
J2245 & s52  &  1 & 0.5785 & $10.9\pm 0.1 $ & $<0.1$ & 509 & 2.54 & 9.5 & $<13.6$ & ... & ... & $<13.8$ & ... & ...\\
J0454 & s53  &  2 & 0.5811 & $11.3\pm 0.1 $ & $<0.5$ & 639 & 1.41 & 10.5 & $<13.4$ & ... & ... & $<13.8$ & ... & ...\\
J0119 & s54  &  1 & 0.5812 & $8.9_{-0.4}^{+0.3}$ & $0.1\pm 0.1 $ & 222 & 2.62 & 8.8 & $<13.3$ & ... & ... & $<13.7$ & ... & ...\\
J2339 & s55  &  1 & 0.5816 & $9.1\pm 0.2 $ & $0.5\pm 0.2 $ & 225 & 2.50 & 8.8 & $<13.3$ & ... & ... & $<13.6$ & ... & ...\\
J0357 & s56  &  4 & 0.5835 & $10.9\pm 0.1 $ & $1.8\pm 0.6 $ & 277 & 1.26 & 10.0 & $13.87_{-0.06}^{+0.06}$ & $26.0_{-4.1}^{+5.0}$ & $79.9_{-4.4}^{+4.7}$ & $<13.4$ & ... & ...\\
J0420 & s57  &  2 & 0.5838 & $9.0\pm 0.1 $ & $0.2\pm 0.1 $ & 79 & 0.90 & 9.7 & $14.45_{-0.03}^{+0.03}$ & $63.4_{-4.6}^{+5.2}$ & $-17.1_{-5.1}^{+5.6}$ & ... & ... & ...\\
J2245 & s58  &  2 & 0.5865 & $10.1\pm 0.1 $ & $0.5\pm 0.3 $ & 115 & 0.98 & 9.6 & $14.11_{-0.06}^{+0.05}$ & $49.3_{-4.0}^{+4.3}$ & $28.6_{-7.5}^{+8.0}$ & $<13.7$ & ... & ...\\
J0333 & s59  &  2 & 0.5872 & $10.8\pm 0.1 $ & $1.1\pm 0.4 $ & 332 & 1.82 & 9.7 & $<13.3$ & ... & ... & $<13.3$ & ... & ...\\
J0114 & s60  &  1 & 0.5887 & $10.5_{-0.2}^{+0.1}$ & $1.1\pm 0.4 $ & 374 & 2.67 & 9.2 & $<13.5$ & ... & ... & $<13.6$ & ... & ...\\
J0154 & s61  &  4 & 0.5890 & $11.2\pm 0.1 $ & $<0.5$ & 967 & 2.37 & 10.2 & $<13.7$ & ... & ... & $<13.4$ & ... & ...\\
J0028 & s62  &  1 & 0.5960 & $10.6\pm 0.1 $ & $0.1\pm 0.1 $ & 287 & 1.92 & 9.4 & $<13.6$ & ... & ... & $<13.3$ & ... & ...\\
J0454 & s63  &  1 & 0.6011 & $10.5\pm 0.1 $ & $<0.1$ & 161 & 1.15 & 9.6 & $<13.2$ & ... & ... & $<13.6$ & ... & ...\\
J0119 & s64  &  9 & 0.6026 & $11.0\pm 0.1 $ & $<0.1$ & 549 & 1.83 & 10.4 & $<13.2$ & ... & ... & $<13.0$ & ... & ...\\
J0357 & s65  &  1 & 0.6076 & $9.7\pm 0.2 $ & $<0.1$ & 220 & 2.14 & 9.0 & $<13.4$ & ... & ... & $<13.8$ & ... & ...\\
J0119 & s66  & 10 & 0.6082 & $11.0\pm 0.1 $ & $<0.1$ & 775 & 2.69 & 10.8 & $13.95_{-0.08}^{+0.07}$ & $25.7_{-5.3}^{+6.0}$ & $98.4_{-5.5}^{+5.7}$ & $<13.5$ & ... & ...\\
J0110 & s67  &  1 & 0.6097 & $9.6\pm 0.1 $ & $1.6\pm 0.4 $ & 266 & 2.65 & 8.9 & $<13.0$ & ... & ... & $<13.3$ & ... & ...\\
J0028 & s68  &  1 & 0.6134 & $11.2\pm 0.1 $ & $0.3\pm 0.2 $ & 959 & 2.38 & 10.2 & $<13.1$ & ... & ... & $<13.4$ & ... & ...\\
J0154 & s69  &  1 & 0.6141 & $10.5\pm 0.1 $ & $<0.1$ & 158 & 1.10 & 9.6 & $<13.5$ & ... & ... & $<13.3$ & ... & ...\\
J0454 & s70  &  1 & 0.6150 & $9.0\pm 0.2 $ & $0.5\pm 0.2 $ & 226 & 2.63 & 8.8 & $<13.4$ & ... & ... & $<13.5$ & ... & ...\\
J2308 & s71  &  4 & 0.6161 & $11.1\pm 0.1 $ & $0.7\pm 0.4 $ & 712 & 2.13 & 10.1 & $<13.7$ & ... & ... & $<13.7$ & ... & ...\\
J2245 & s72  &  5 & 0.6179 & $10.6\pm 0.1 $ & $<0.1$ & 173 & 1.17 & 10.1 & $<13.9$ & ... & ... & $14.00_{-0.10}^{+0.09}$ & $26.3_{-7.9}^{+10.2}$ & $63.1_{-6.5}^{+6.6}$\\
J2135 & s73  & 13 & 0.6221 & $11.0\pm 0.1 $ & $0.7\pm 0.2 $ & 128 & 0.46 & 10.9 & $14.84_{-0.12}^{+0.32}$ & $24.6_{-10.4}^{+10.4}$ & $-28.3_{-2.5}^{+4.0}$ & ... & ... & ...\\
J0357 & s74  &  1 & 0.6258 & $9.1\pm 0.1 $ & $1.9\pm 0.2 $ & 153 & 1.73 & 9.0 & $<13.6$ & ... & ... & $<13.5$ & ... & ...\\
J0111 & s75  &  2 & 0.6300 & $11.0\pm 0.1 $ & $<0.1$ & 501 & 1.93 & 10.0 & $<13.1$ & ... & ... & ... & ... & ...\\
J2308 & s76  &  1 & 0.6352 & $8.6_{-0.5}^{+0.3}$ & $0.1\pm 0.1 $ & 214 & 2.79 & 8.7 & $<13.3$ & ... & ... & $<13.5$ & ... & ...\\
J2245 & s77  &  5 & 0.6386 & $10.2\pm 0.1 $ & $4.9\pm 0.5 $ & 184 & 1.53 & 9.9 & $<13.4$ & ... & ... & $<13.6$ & ... & ...\\
J0111 & s78  &  4 & 0.6483 & $10.9\pm 0.1 $ & $<0.1$ & 59 & 0.29 & 10.6 & $<13.3$ & ... & ... & ... & ... & ...\\
J2245 & s79  &  2 & 0.6610 & $11.2\pm 0.1 $ & $<0.1$ & 556 & 1.34 & 10.5 & $<13.6$ & ... & ... & $<13.6$ & ... & ...\\
J0454 & s80  &  3 & 0.6617 & $9.5\pm 0.1 $ & $2.4\pm 0.4 $ & 183 & 1.88 & 9.4 & $13.80_{-0.07}^{+0.08}$ & $25.0_{-5.1}^{+6.0}$ & $-151.0_{-5.7}^{+5.4}$ & $13.74_{-0.11}^{+0.09}$ & $<32.2$ & $-144.9_{-6.5}^{+5.4}$\\
J0420 & s81  &  2 & 0.6641 & $10.1\pm 0.1 $ & $1.0\pm 0.2 $ & 121 & 1.04 & 9.6 & $13.89_{-0.11}^{+0.10}$ & $25.0_{-8.1}^{+24.7}$ & $2.2_{-6.4}^{+4.6}$ & $<13.7$ & ... & ...\\
J2308 & s82  &  2 & 0.6643 & $8.1\pm 0.3 $ & $0.1\pm 0.1 $ & 78 & 1.23 & 9.3 & $<13.2$ & ... & ... & $<14.1$ & ... & ...\\
J0248 & s83  &  1 & 0.6645 & $7.7\pm 0.1 $ & $0.3\pm 0.2 $ & 46 & 0.83 & 8.9 & $<13.1$ & ... & ... & $<13.2$ & ... & ...\\
J0114 & s84  &  2 & 0.6652 & $10.5\pm 0.2 $ & $0.3\pm 0.2 $ & 393 & 2.83 & 9.3 & $<14.0$ & ... & ... & $<13.7$ & ... & ...\\
J2135 & s85  &  1 & 0.6718 & $9.8\pm 0.1 $ & $0.5\pm 0.2 $ & 313 & 2.96 & 9.0 & $<13.1$ & ... & ... & ... & ... & ...\\
J2339 & s86  &  2 & 0.6724 & $9.0\pm 0.1 $ & $1.7\pm 0.3 $ & 180 & 2.15 & 9.3 & $<13.9$ & ... & ... & $<13.3$ & ... & ...\\
J2135 & s87  &  1 & 0.6758 & $7.8\pm 0.3 $ & $0.1\pm 0.1 $ & 33 & 0.59 & 9.1 & $13.70_{-0.13}^{+0.10}$ & $27.1_{-10.9}^{+14.2}$ & $0.3_{-9.0}^{+7.7}$ & ... & ... & ...\\
J2339 & s88  &  1 & 0.6779 & $8.2\pm 0.2 $ & $0.3\pm 0.2 $ & 86 & 1.33 & 8.9 & $<13.4$ & ... & ... & $<13.4$ & ... & ...\\
J0357 & s89  &  3 & 0.6802 & $8.4\pm 0.5 $ & $0.1\pm 0.1 $ & 191 & 2.63 & 10.7 & $<13.1$ & ... & ... & $<13.4$ & ... & ...\\
J0333 & s90  &  1 & 0.6802 & $8.3_{-0.4}^{+0.3}$ & $0.1\pm 0.1 $ & 158 & 2.25 & 8.7 & $<13.2$ & ... & ... & $<13.6$ & ... & ...\\
J0110 & s91  &  2 & 0.6803 & $9.6\pm 0.1 $ & $1.7\pm 0.4 $ & 147 & 1.49 & 9.4 & $13.59_{-0.11}^{+0.09}$ & $31.7_{-7.0}^{+7.7}$ & $-40.8_{-8.1}^{+8.8}$ & $<13.1$ & ... & ...\\
J0154 & s92  &  1 & 0.6824 & $9.4\pm 0.2 $ & $1.1\pm 0.3 $ & 189 & 1.99 & 9.0 & $<13.3$ & ... & ... & $<13.4$ & ... & ...\\
J0248 & s93  &  1 & 0.6885 & $8.7\pm 0.2 $ & $<0.1$ & 195 & 2.52 & 8.8 & $<13.3$ & ... & ... & $<13.1$ & ... & ...\\
J0111 & s94  &  1 & 0.6898 & $10.9\pm 0.1 $ & $<0.2$ & 372 & 1.66 & 9.9 & $<13.0$ & ... & ... & ... & ... & ...\\
J0357 & s95  &  5 & 0.6922 & $8.7_{-0.5}^{+0.3}$ & $0.1\pm 0.1 $ & 98 & 1.27 & 10.9 & $<13.9$ & ... & ... & $<13.6$ & ... & ...\\
J0454 & s96  &  2 & 0.6938 & $10.6\pm 0.1 $ & $<0.2$ & 392 & 2.71 & 9.4 & $<13.4$ & ... & ... & $<13.3$ & ... & ...\\
J0333 & s97  &  1 & 0.6945 & $10.5\pm 0.1 $ & $0.2\pm 0.2 $ & 307 & 2.21 & 9.3 & $14.28_{-0.05}^{+0.05}$ & $15.2_{-2.3}^{+2.7}$ & $63.1_{-2.1}^{+2.0}$ & $<13.6$ & ... & ...\\
J0357 & s98  &  4 & 0.7043 & $8.9\pm 0.2 $ & $0.4\pm 0.2 $ & 188 & 2.30 & 9.7 & $<13.0$ & ... & ... & $<13.4$ & ... & ...\\
J0420 & s99  &  2 & 0.7046 & $10.8\pm 0.1 $ & $<0.1$ & 295 & 1.60 & 9.8 & $14.06_{-0.10}^{+0.19}$ & $<24.6$ & $-66.5_{-3.4}^{+5.3}$ & $<13.6$ & ... & ...\\
J2135 & s100 &  1 & 0.7046 & $10.2_{-0.2}^{+0.1}$ & $0.3\pm 0.2 $ & 346 & 2.86 & 9.1 & $<13.6$ & ... & ... & ... & ... & ...\\
J0248 & s101 &  4 & 0.7062 & $9.7\pm 0.1 $ & $2.0\pm 0.4 $ & 227 & 2.25 & 10.3 & $13.91_{-0.10}^{+0.09}$ & $<44.0$ & $140.6_{-10.0}^{+7.9}$ & $<13.6$ & ... & ...\\
J0110 & s102 &  3 & 0.7092 & $11.3\pm 0.1 $ & $0.6\pm 0.3 $ & 1186 & 2.83 & 10.3 & $<13.2$ & ... & ... & $<13.6$ & ... & ...\\
J0119 & s103 &  6 & 0.7182 & $11.4\pm 0.1 $ & $<0.2$ & 674 & 1.38 & 10.9 & $<13.2$ & ... & ... & $<13.5$ & ... & ...\\
\enddata
\end{deluxetable*}
\end{longrotatetable}



\end{document}